\pdfoutput=1

\documentclass[12pt,preprint]{aastex}

\usepackage{graphicx}
\usepackage{amssymb}
\usepackage{natbib}
\usepackage{fixltx2e}
\usepackage{rotating}
\usepackage{natbib}
\usepackage{amsmath}
\usepackage{pdfpages}
\usepackage{definitions} 
\begin{document}

\title{{\it Herschel} Galactic plane survey of \NII\ fine structure emission}

\author{Paul F. Goldsmith\altaffilmark{1} 
\email{Paul.F.Goldsmith@jpl.nasa.gov}}
  
\author{Umut A. Y{\i}ld{\i}z\altaffilmark{1}} 
\author{William D. Langer\altaffilmark{1}} 
\author{Jorge L. Pineda\altaffilmark{1}}

\altaffiltext{1}{Jet Propulsion Laboratory, California Institute of Technology, 4800 Oak Grove Drive, Pasadena CA, 91109, USA}

\date{Accepted for Publication in ApJ, October 2015}

\begin{abstract}
We present the first large scale high angular resolution survey of ionized nitrogen in the Galactic Plane through emission of its two fine structure transitions (\NII) at 122~\um\ and 205~\um. The observations were largely obtained with the PACS instrument onboard the \herschellong. The lines--of--sight were in the Galactic plane,  following those of the \textit{Herschel} OTKP project GOT C+.
Both lines are reliably detected at the 10$^{-8}$ -- 10$^{-7}$ \Wmtwosr\ level over the range --60$^o$ $\leq$ {\it l} $\leq$ 60$\degr$.  The $rms$ of the intensity among the 25 PACS spaxels of a given pointing is typically less than one third of the mean intensity, showing that the emission is extended.  \NII\ is produced in gas in which hydrogen is ionized, and collisional excitation is by electrons.  
The ratio of the two fine structure transitions provides a direct measurement of the electron density, yielding $n(e)$ largely in the range 10 to 50 \cmthree\ with an average value of 29 \cmthree\ and  \nplus\ column densities 10$^{16}$ to 10$^{17}$ \cmtwo.  \NII\ emission is highly correlated with that of \CII, and we calculate that  between 1/3 and 1/2 of the \CII\ emission is associated with the ionized gas.  The relatively high electron densities indicate that the source of the \NII\ emission is not the Warm Ionized Medium (WIM), which has electron densities more than 100 times smaller.   Possible origins of the observed \NII\ include the ionized surfaces of dense atomic and molecular clouds, the extended low density envelopes of \HII\ regions, and low--filling factor high--density fluctuations of the WIM.

\end{abstract}


\section{INTRODUCTION}

The atoms and molecules in the interstellar medium (ISM) play an important role in the formation and evolution of the Galaxy, providing critical cooling of various phases.   In addition, far infrared, submillimeter, and millimeter wavelength atomic and ionic fine structure and molecular rotational lines are powerful tracers of star formation on both Galactic and extragalactic scales. Although rotational transitions of CO trace cool--to--moderately warm molecular gas, ionized carbon produces the strongest far-infrared line (\CII), which arise from almost all reasonably warm ($T$$>$40~K) UV suffused portions of the ISM. However, \CII\ alone cannot distinguish highly ionized gas from weakly ionized, neutral, or even moderate-extinction molecular gas.  The ionization potential of nitrogen (14.5 eV) is greater than that of hydrogen (13.6 eV), therefore the ionized nitrogen \NII\ lines reflect the effect of UV photons emitted by massive young stars, with possible enhancement from X-ray photoionization.  Since \nplus\  is produced only in fully ionized regions (\HII\ regions as well as diffuse ionized gas), \NII\ emission can be used to separate the fully and weakly ionized regions.  

The 158~\um\ line of ionized carbon \CII\ is the strongest single far-IR emission feature from the interstellar medium and the most important coolant of gas in which hydrogen is neutral \citep{Bennett94, Fixsen99}. It is a key determinant of the evolution of largely atomic regions into denser, cooler, molecular clouds in which new stars are formed, and is widely used as a tracer of star formation in the Milky Way and other galaxies \citep{Stacey91, Malhotra01, Contursi02, Stacey10, Pineda14}. There is an ongoing question about the origin of the \CII\ emission, which has been asserted to come from the extended low density warm interstellar medium \citep{Heiles01, Abel06}, but has more generally been associated with  photon dominated regions (PDRs) intimately associated with massive, young stars \citep{Crawford85, Stacey83, Stacey85, Stacey91, Stacey10, Pineda14, Velu14} and also the ``CO--Dark H$_2$ gas'' \citep{Pineda13, Langer14}.  Determining the distribution and characteristics of \NII\ emission can provide valuable information as it is expected to originate exclusively in ionized regions. 

In this paper, we report the results of the first large--scale Galactic survey of the 122~\um\ and 205~\um\ \NII\ line observations comprising observations of 149 positions in the Galactic Plane.  These data were obtained using the PACS instrument onboard the \herschellong. 
We also present velocity resolved \NII\ 205~\um\ spectra at 10 positions obtained with the HIFI instrument. 
In Section \ref{sec:obs}, we discuss the observational and data reduction procedures used, and present the results in Section \ref{sec:results}.
In Section \ref{sec:excitation}, we discuss collisional excitation of \NII\ and the procedures used to derive the electron densities and \NII\ column densities.
In Sections \ref{electron_dens} and \ref{sec:discussion} we present and discuss the derived electron densities and \nplus\ column densities, compare \NII\ with \CII\ emission, and discuss different possible sources of the observed \NII\ emission.  We summarize our results in Section \ref{sec:conclusions}.

\section{OBSERVATIONS}
\label{sec:obs}

\subsection{[N\,{\sc ii}] Observations}

Observations of both fine structure transitions of \NII, at 121.898~\um\ (2459.371425~GHz) and 205.178~\um\ (1461.134~GHz) \citep{brown94}, were carried out with the PACS spectrometer \citep{Poglitsch10} on {\it Herschel} \citep{Pilbratt10}.  At 10 selected positions, the lower frequency transition was also observed with the high spectral resolution \hifilong\ \citep[HIFI;][]{degraauw10}.   Based on the expected relative intensities of the \NII\ and \CII\ lines \citep{Bennett94}, observations were largely restricted to the inner galaxy, with both lines detected only in the range -57.4$\degr$ $\leq$ {\it l} $\leq$ 78.1$\degr$.

\begin{figure}[!t]
    \centering
	\includegraphics[scale=0.4]{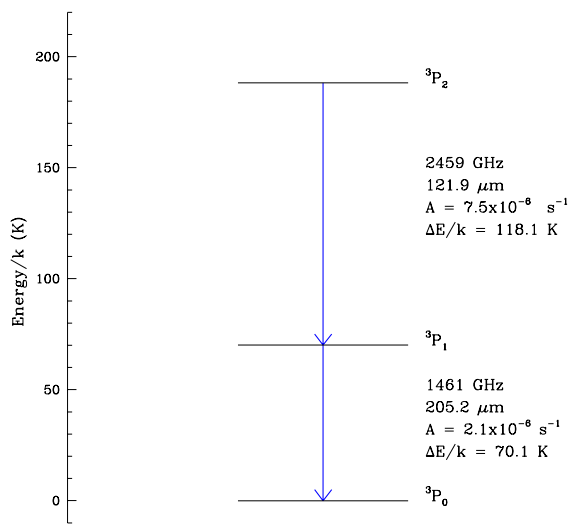}
      \caption{Energy level diagram of the \NII\ fine structure transitions in the electronic ground state of N$^+$ including the frequency, wavelength, and spontaneous decay rate for each transition.}
    \label{fig:NIItransitions}
\end{figure}

Table~\ref{tbl:overviewlines} gives the upper level energies, Einstein $A$--coefficients \citep{Galavis97}, and wavelengths of the two \NII\ transitions, as well as the single fine structure transition of \CII.
Both transitions of  \NII\ are split by hyperfine interactions; the splittings of the \thPtwo\ -- \thPone\  transition correspond to as much as 50 \kms\ \citep{brown94}, but as this line was observed only with PACS with a velocity resolution several times worse than this splitting, we have not made any correction for the hyperfine structure.   The splittings of the \thPone\ -- \thPzero\ transition correspond to less than 1.5 \kms\ and thus in principle could be resolved with HIFI, but the observed line widths are considerably greater, so we have ignored the splittings in analyzing the present data.
In Fig.~\ref{fig:NIItransitions}, we show the term scheme of the fine structure transitions in the electronic ground state of ionized nitrogen. \\

\begin{table}[!t]
\caption{Characteristics of the fine structure transitions of \NII\ and \CII.}
\begin{center}
\begin{tabular}{rcrcccc}
\hline \hline
Ion & Levels & $E_\mathrm{u}/k$ & $A_{\rm ul}$ & Wavelength & Frequency & $\Delta E/k$ \\
 &   &  [K] & [\ss] & [$\mu$m]  & [GHz] & [K]\\
\hline
\noalign{\smallskip}
N$^+$ & \thPtwo\ -- \thPone\        & 188.1    & 7.5$\times$10$^{-6}$   & 121.898 & 2459.374\tablenotemark{1} & 118.0\phantom{0}\\
N$^+$ & \thPone\ -- \thPzero\       & 70.1      & 2.1$\times$10$^{-6}$   & 205.178 & 1461.131\tablenotemark{1} & 70.1   \\
C$^+$ & $^{2}\rm{P}_{3/2}$ -- $^{2}\rm{P}_{1/2}$ & 91.2      & 2.3$\times$10$^{-6}$   & 157.740 & 1900.537 & 91.2   \\
\noalign{\smallskip}
\hline
\end{tabular}
\tablenotetext{1}{Fitted values from Table 4 of \citet{brown94} weighted by relative hyperfine intensities}
\end{center}
\label{tbl:overviewlines}
\end{table}

\begin{figure}[tb]
    \centering
    \includegraphics[scale=0.5]{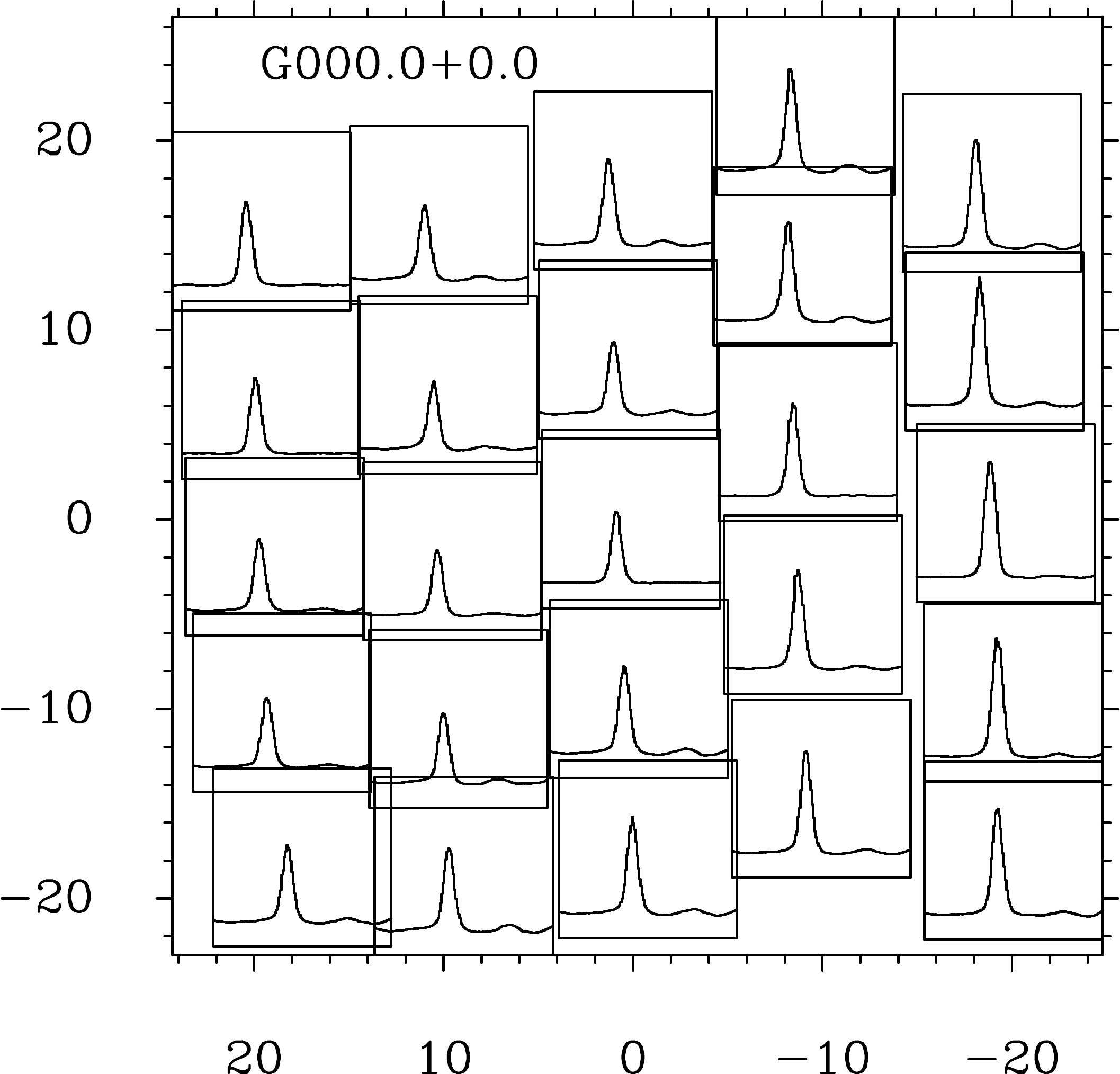}
     \caption{\small An example of the data in a single frequency band of a PACS observation.  We denote Galactic coordinates by their Galactic longitude+latitude. Here, the central spaxel of the array is pointed towards the Galactic Center, G000+0.0.  The offsets of the PACS pixels are given in seconds of arc. The \NII\ 122~\um\ line is clearly detected in each of the 25 spaxels, but the profile seen is the instrumental response function rather than the intrinsic line profile due to the relatively low resolution (R $\simeq$ 1000) of PACS.    The velocity range covered is approximately 5000 \kms\ and the maximum flux density is approximately 80 Jy/beam.}
     \label{fig:PACSfootprint}
\end{figure}

{\bf PACS:} The PACS data comprise 149 line-of-sight (LoS) positions observed at a Galactic latitude of $b$=0$\degr$ with spacings of $\sim$0.9$\degr$ near the Galactic center and $\sim$4$\degr$ up to 13$\degr$ in the anticenter direction.  The LoS positions were selected from the GOT C+ survey of the \CII\ 158~\um\ transition \citep{Langer10, Pineda13}. The off positions were at Galactic latitudes of $b$ = $\pm$2$\degr$ at the longitude of each of the Galactic plane pointing directions.  The \NII\ observations were conducted between August 2012 and March 2013.  Each LoS was observed with an integration time of 897 seconds including the off--source observation.  The PACS spectrometer has an array of 5$\times$5 spatial pixels (spaxels), each with a resolving power of $\simeq$ 1000 at 122~\um\ and $\simeq$ 2000 at 205~\um.  The nominal pointing accuracy of the telescope is 2$\arcsec$ \rms.
The {\it Herschel} telescope with PACS has a FWHM beam width of $\simeq$10$\arcsec$ at 122~\um\ and 15\arcsec\ at 205~\um.  The actual beam profile is not highly Gaussian at low levels relative to the on--axis response (especially at the shorter wavelength) due to the 9.4\arcsec\ x 9.4\arcsec\ size of the detector pixels and wavefront errors \citep{Poglitsch10}. The two  \NII\ transitions were observed  simultaneously using the ``PACS range-scan mode''. 
Figure~\ref{fig:PACSfootprint} gives an example of a PACS footprint obtained when pointing towards the Galactic Center. 

\begin{figure}[tb]
    \centering
	\includegraphics[scale=0.5]{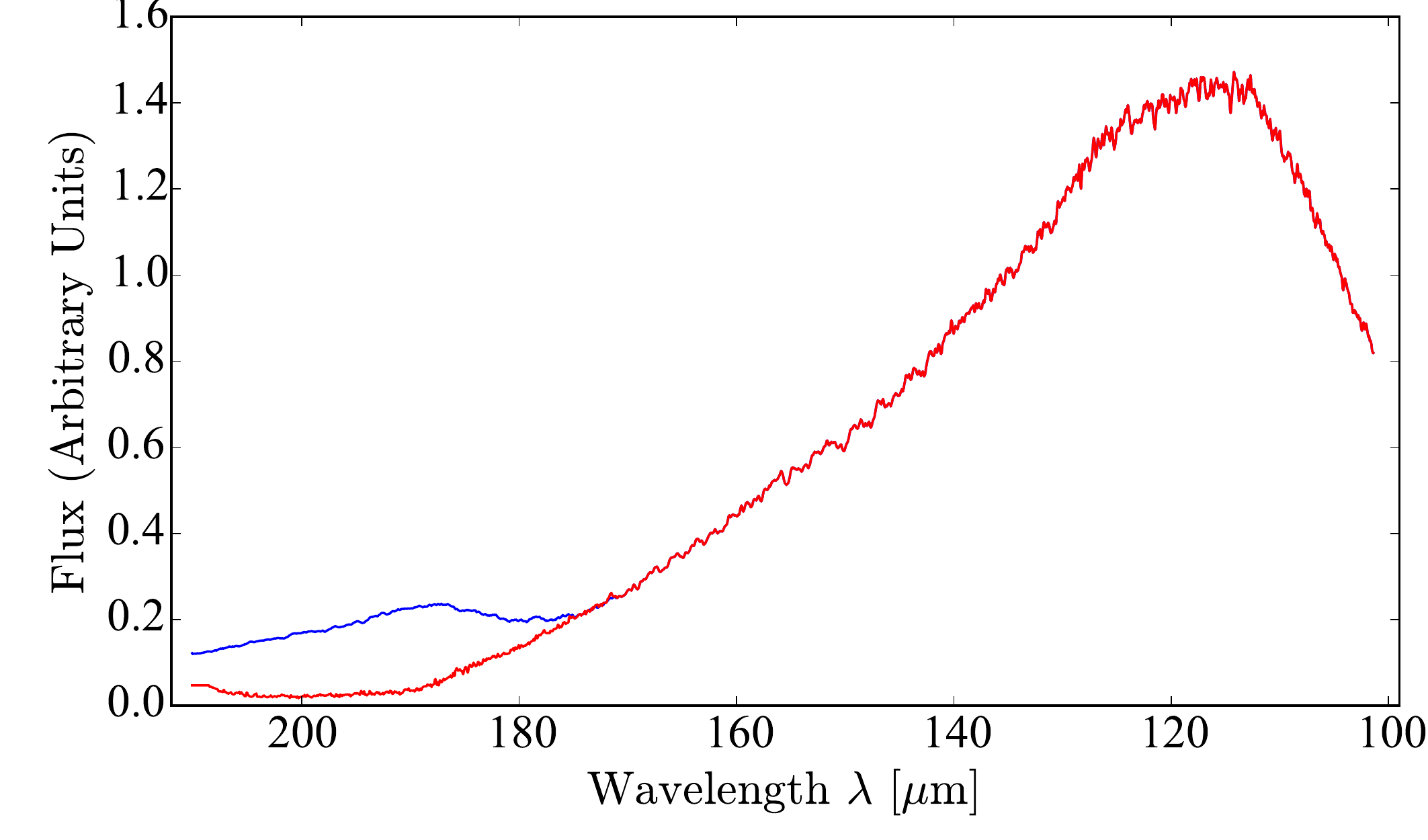}
     \caption{Relative response of PACS showing the effect of the red filter leak.  The left hand side corresponds to longer wavelengths and the right hand side to shorter wavelengths.  The blue curve shows the response of the spectrometer including the leak, while the red curve is the correct relative spectral response function.  The two responses have major differences in the 190 \um\ to 210 \um\ range.}
     \label{fig:redleak}
\end{figure}

 We used \hipelong\ (HIPE) version 10.3 \citep{OttS10} for the PACS data reduction.
For the reduction of the 122~\um\ line data, PACS calibration tree version \#65 was used. However, for the 205~\um\ line spectra, an earlier version of calibration tree, version \#49 was employed. 
The reason is that the 205~\um\ line lies in the region affected by the   ``red filter leak'' or ``order light leak\footnote{Steve Lord, private communication.}", shown in Fig.~\ref{fig:redleak}.
This leak is due to energy from the second order grating response at $\simeq$ one half the nominal wavelength being added to the signal at the observing wavelength, and thus contaminating the 190--220~\um\ range.

The leakage of radiation from shorter wavelengths is registered as signal at wavelengths between 190~\um\ and 210~\um\ range  and results in an incorrectly high relative spectral response function (RSRF) in this wavelength range.  The erroneous RSRF results in an underestimate of the strength of spectral lines.
The corrected RSRF increased the line flux by a factor $\simeq$ 3.  Fortunately, we had observations of the 205~\um\ \NII\ line at 10 positions with HIFI, and comparison of the intensities derived from PACS and HIFI data showed that the corrected RSRF yielded intensities substantially consistent with the HIFI data, as discussed further below.
 
The HIPE pipeline produces fluxes (\Wm2), which we then converted to intensities (\Wmtwosr) by dividing by the PACS pixel solid angle $\Omega_{\rm pix}$, equivalent to a flat--topped 9.4$\arcsec$$\times$9.4$\arcsec$ beam with solid angle  2.35$\times$10$^{-9}$ sr, as appropriate for a uniform source filling the beam. Integrated intensities $I$, in units of \Wmtwosr, are used throughout this paper.  We present the PACS data in Tables 2 \& 3, which give for each line of sight a label (column 2), \herschel\ OBSID (column 3), the \NII\ 122~\um\ intensity and its \rms\ error averaged over all 25 spaxels (column 5), the standard deviation of the intensity\footnote{The variation among the spaxels of a pointing is discussed in Section \ref{variation}.}, $\sigma$ (column 6), and the 205~\um\ intensity and its \rms\ error similarly averaged (column 8). 

\begin{figure}[tb]
	\centering
	\includegraphics[scale=0.5]{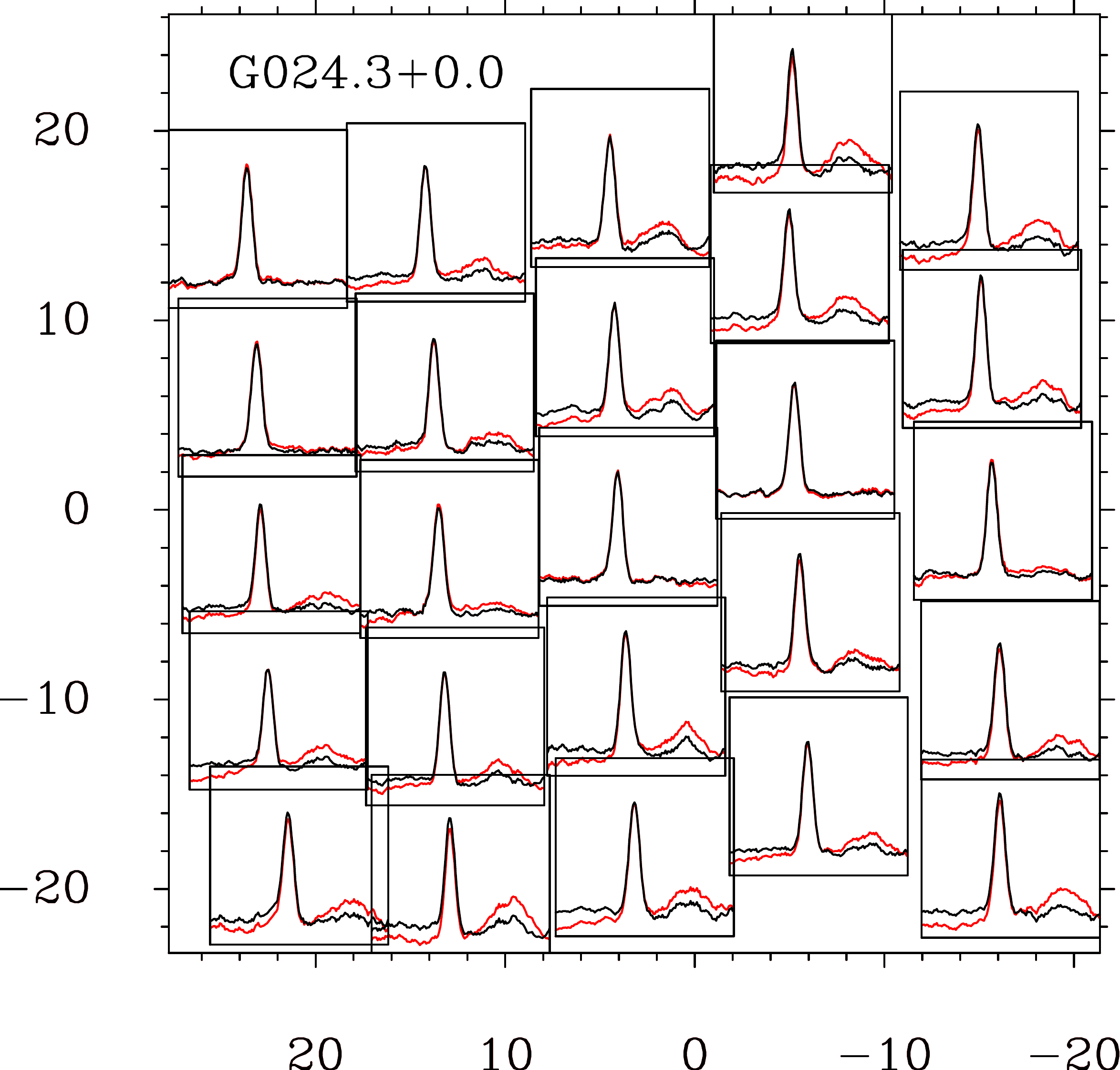}
	\caption{Illustration of effect of polynomial baseline removal from PACS spectra towards G024.3+0.0.  A DC offset had previously been removed to yield the 25 spectra displayed as the red lines in this 122 \um\ observation.  The higher--order polynomial (spectra shown in black after baseline fitting and removal) modestly reduces any residual offset in the vicinity of the line as well as reducing the amplitude of the feature to the right of the \NII\ emission.}
	\label{fig:PACS-baselines}
\end{figure}

Instrumental baseline removal was non--trivial for these data.  Due to the high variability in the baseline, the wavelength range including the target line was carefully masked for each of the observations, and a baseline including up to a fifth order polynomial was fitted and removed from the data. 
Figure \ref{fig:PACS-baselines} shows the effect of baseline removal on the PACS data.  In the present example, a zeroth order baseline has previously been fitted to remove a DC offset which was much larger than the signal.  The higher-order polynomial removes any residual offset in the vicinity of the line as well as slightly reducing the amplitude of the feature to the right of the emission.  Using a high order polynomial is safe in this case because the spectral resolution of PACS does not allow the possibility of resolving Galactic lines and we calculated only the (integrated) flux for each spaxel. The \rms\ of the observations is typically $\sim$2$\times$10$^{-10}$ \Wmtwosr\ for the 122~\um\ line and $\sim$6$\times$10$^{-10}$ \Wmtwosr\  for the 205~\um\ line.

\includepdf[pages={1}]{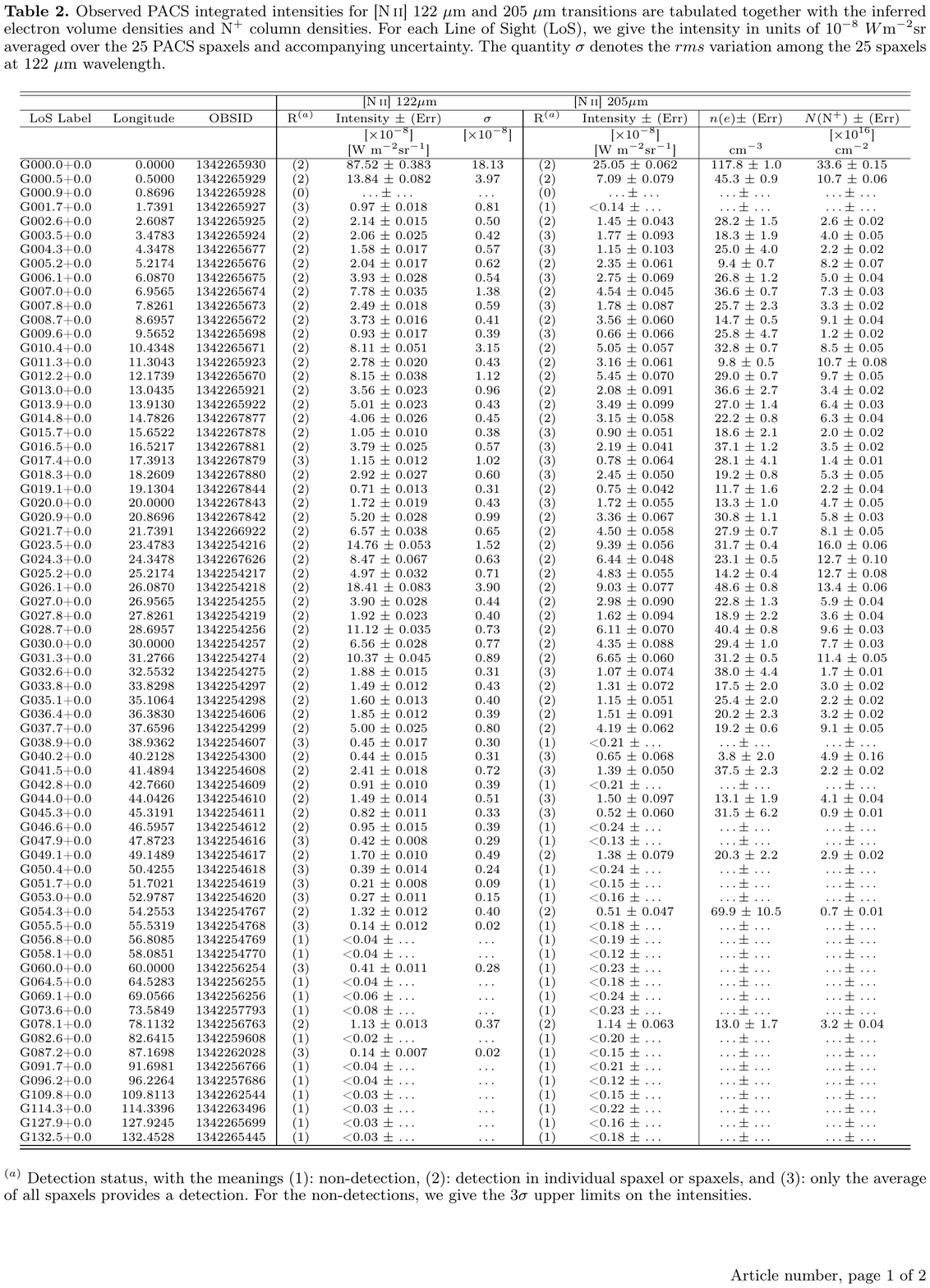}
\includepdf[pages={2}]{appendixtable.pdf}

%

{\bf HIFI:} We observed 10 LoS directions using HIFI 
 in \NII\ 205~\um\ (band 6a) and \CII\ 158~\um\ \citep[band 7b]{Pineda13}.   The HIPE pipeline produced data in terms of  antenna temperature (\Tas), which were then converted to main-beam brightness temperatures $T_{\rm MB} = T_{\rm A}^{*}/ \eta_{\rm MB}$ by using a beam  efficiency $\eta_{\rm MB}$ of 0.60 \citep[updates in HIFI Beam Release Notes Oct 2014]{Roelfsema12}.

The data were smoothed to a velocity resolution of 1~\kms\ in order to increase the signal to noise ratio.   
The \NII\ 205 \um\ HIFI data show a prominent artifact at the lower end of the frequency scale. We removed these artifacts by setting the parameter {\it width} to be 10 MHz instead of 30 MHz in the {\tt mkOffSmooth} task (note that these settings will become standard in products generated from HIPE 14 onward; Herschel Science Center, private communicaton).

\begin{figure}[!t]
	\centering
		\includegraphics[scale=0.5]{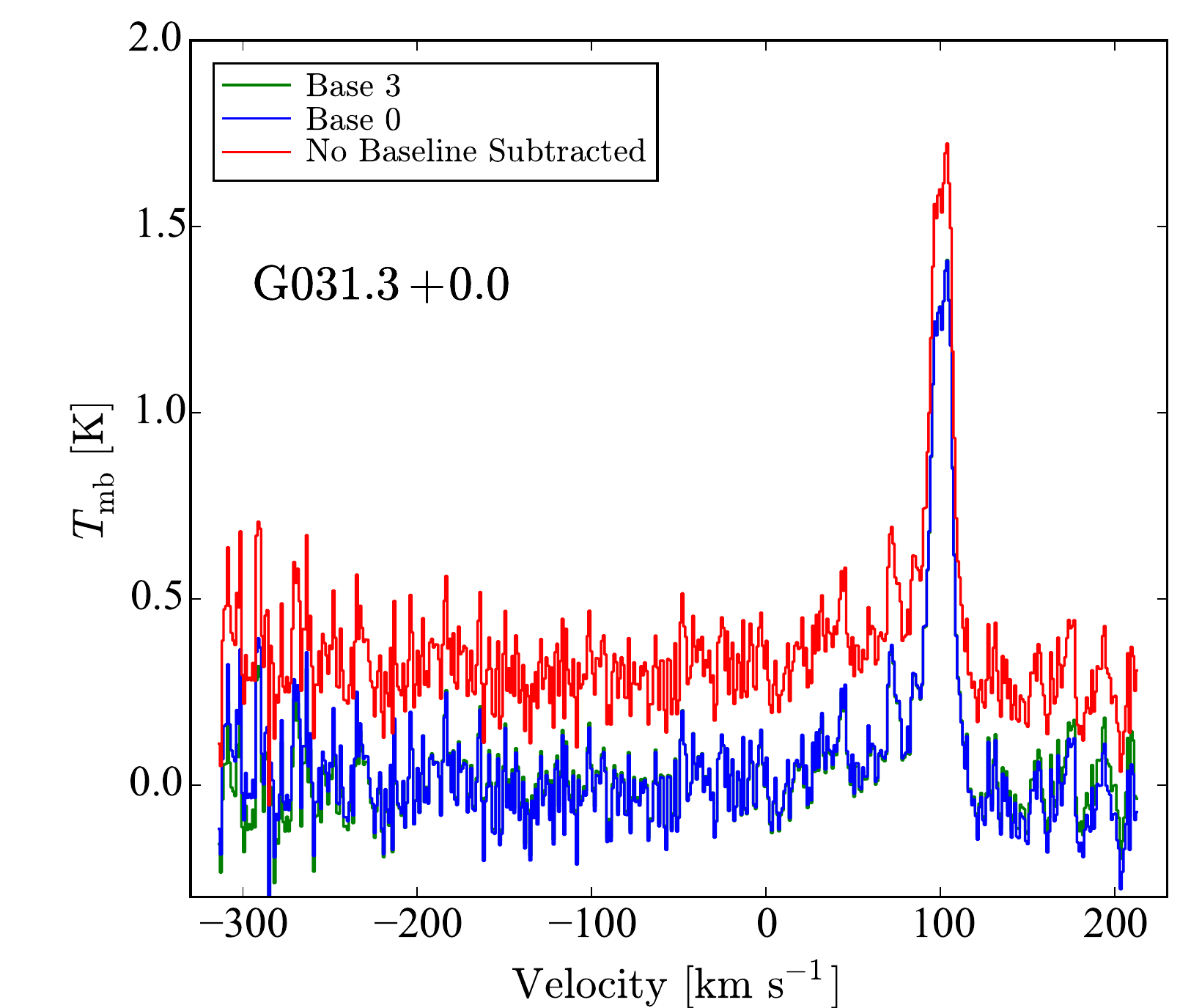}
		\caption{HIFI \NII\ spectrum before baseline removal (with positive continuum offset; shown in black), with continuum offset removed (lower spectrum; shown in blue), and with third order baseline removed (lower spectrum; shown in red).  There is only a minimal difference between the spectra with zeroth and third order baselines removed in this case.}
		\label{NII_G031.3_before_after}
\end{figure}

A polynomial, of order between 0 and 3, was fitted and removed to produce flat spectral baselines.  Figure \ref{NII_G031.3_before_after} shows a typical spectrum before and after baseline removal.  In this case, a linear baseline is all that is required.  
The FWHM width of the {\it Herschel} HIFI beam is 15.7$\arcsec$ at the 1461~GHz frequency (\NII) and 12.1$\arcsec$ at 1900~GHz (\CII).
There is a slight difference between the pointing directions of the H and V polarizations in HIFI, but this difference, $(\delta \rm{H}, \delta \rm{V})$ = \citep[+0$\farcs$7, +0$\farcs$3;][]{Roelfsema12} for Band 6a, is small enough that it can be neglected relative to the beam size.

An on+off integration time of 7041 s ($\sim$2 hours) per pointing was used to obtain the HIFI data.  This yielded an rms antenna temperature $\sigma(T_A)$ $\simeq$ 0.1 K in a 1~\kms\ channel.  For the diffraction--limited {\it Herschel} HIFI beam, this rms is equivalent to an uncertainty in the intensity $\sigma(I)$ $\simeq$ 3.2$\times$10$^{-8}$ \Wmtwosr.  

\begin{figure}[!t]
    \centering
    \includegraphics[scale=0.6]{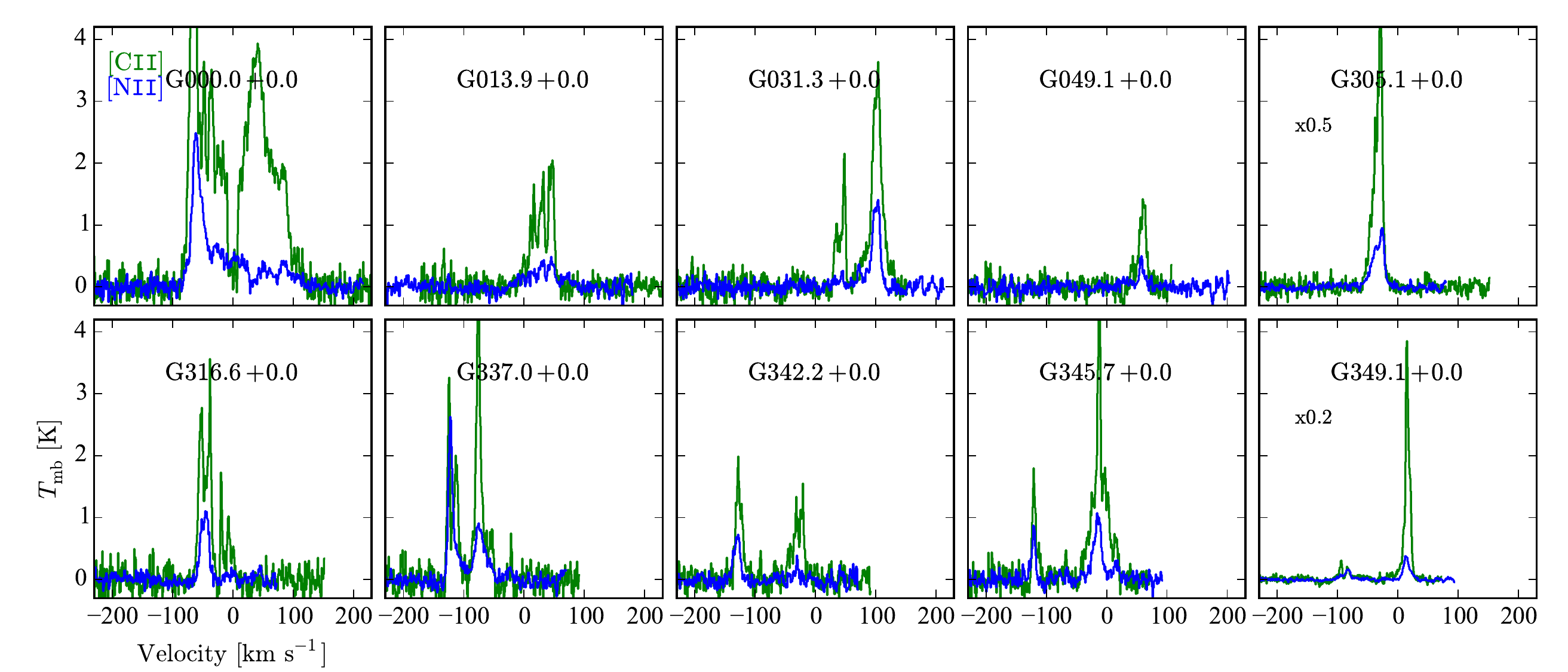}
    \caption{ Spectra of \NII\ 205~\um\ obtained with the  {\it Herschel} HIFI instrument are shown by the blue lines.  The \CII\ 158~\um\ spectra at the same Galactic coordinates (given for each pointing direction) from the GOT C+ project are shown by the green lines.   A scaling factor for the \CII\  data is indicated with the two lines of sight for which it was used to display the relatively strong \CII\  data.}
    \label{fig:HIFINIICIIspectra}
\end{figure}

We show the 10 HIFI \NII\ 205~\um\ spectra in Fig.~\ref{fig:HIFINIICIIspectra}, in which we also overlay the \CII\ 158~\um\ spectra obtained by the GOT C+ project.  Table~\ref{HIFI_intensities}  gives the intensities  and uncertainties  of the \NII\ lines observed with HIFI, obtained by integrating over velocity and using the formula appropriate for a diffraction--limited beam $I$ (\Wmtwosr) = 3.13$\times$10$^8 \int T_A dv$ (K kms$^{-1}$).  The intensities and uncertainties of the \CII\ 158~\um\ line from the GOT C+ project are also included.  A number of interesting characteristics emerge from this relatively small spectrally--resolved FIR line data set.  First, we note there is never any clear detection of \NII\ emission at a velocity that does not have \CII\ emission.  What we see quite generally are spectral features in which {\it both} emission lines are present, albeit with highly variable intensity ratio. Second, we see that there is typically a modest number, 2 to 6, of distinct velocity components in a given spectrum, which of course would not be resolved by PACS.  We shall return to these points in our subsequent discussion of the origin of the \NII\ emission.

\begin{table}[!th]
\renewcommand\thetable{4} 
\caption{\label{HIFI_intensities}  205~\um\ \NII\  and 158~\um\ \CII\ intensities observed with the {\it Herschel} HIFI instrument}
\small  
\begin{center}
\begin{tabular}{c c c c c c c }
\hline \hline
& & & \multicolumn{2}{c}{[N\,{\sc ii}] 205$\mu$m} & \multicolumn{2}{c}{[C\,{\sc ii}] 158$\mu$m}  \\ \hline
LoS Label & Longitude        & OBSID\tablenotemark{a}      & Intensity & Uncertainty &  Intensity & Uncertainty   \\ 
\hline
     &     &       & [W m$^{-2}$sr$^{-1}$] & [W m$^{-2}$sr$^{-1}$] &  [W m$^{-2}$sr$^{-1}$] & [W m$^{-2}$sr$^{-1}$]   \\ 
\hline
G000.0+0.0 & 000.0000 & 1342268193 & 3.14$\times$10$^{-7}$ & 4.64$\times$10$^{-9}$ & 2.57$\times$10$^{-6}$ & 1.60$\times$10$^{-8}$  \\ 
G013.9+0.0 & 13.9130 & 1342268194 & 4.92$\times$10$^{-8}$ & 2.76$\times$10$^{-9}$ & 3.96$\times$10$^{-7}$ & 4.87$\times$10$^{-9}$  \\ 
G031.3+0.0 & 31.2766 & 1342254395 & 8.82$\times$10$^{-8}$ & 3.31$\times$10$^{-9}$ & 5.92$\times$10$^{-7}$ & 6.54$\times$10$^{-9}$  \\ 
G049.1+0.0 & 49.1489 & 1342255764 & 2.05$\times$10$^{-8}$ & 2.87$\times$10$^{-9}$ & 1.54$\times$10$^{-7}$ & 5.34$\times$10$^{-9}$  \\ 
G305.1+0.0 & 305.1060 & 1342262675 & 1.16$\times$10$^{-7}$ & 2.13$\times$10$^{-9}$ & 7.83$\times$10$^{-7}$ & 5.33$\times$10$^{-9}$  \\ 
G316.6+0.0 & 316.5960 & 1342262676 & 5.73$\times$10$^{-8}$ & 2.97$\times$10$^{-9}$ & 4.60$\times$10$^{-7}$ & 8.01$\times$10$^{-9}$  \\ 
G337.0+0.0 & 336.9570 & 1342263178 & 1.57$\times$10$^{-7}$ & 3.59$\times$10$^{-9}$ & 7.56$\times$10$^{-7}$ & 8.30$\times$10$^{-9}$  \\ 
G342.2+0.0 & 342.1740 & 1342263177 & 4.41$\times$10$^{-8}$ & 3.61$\times$10$^{-9}$ & 2.75$\times$10$^{-7}$ & 8.18$\times$10$^{-9}$  \\ 
G345.7+0.0 & 345.6520 & 1342263176 & 8.39$\times$10$^{-8}$ & 3.76$\times$10$^{-9}$ & 5.71$\times$10$^{-7}$ & 8.68$\times$10$^{-9}$  \\ 
G349.1+0.0 & 349.1300 & 1342267948 & 1.14$\times$10$^{-7}$ & 3.61$\times$10$^{-9}$ & 1.20$\times$10$^{-6}$ & 1.31$\times$10$^{-8}$  \\ 
\hline
\hline
\end{tabular}
\tablenotetext{a} {Herschel Observation Identification Number.}
\end{center}
\label{tbl:HIFI_intensities}
\end{table}

\begin{figure}[!t]
    \centering
    \includegraphics[width=0.9\columnwidth]{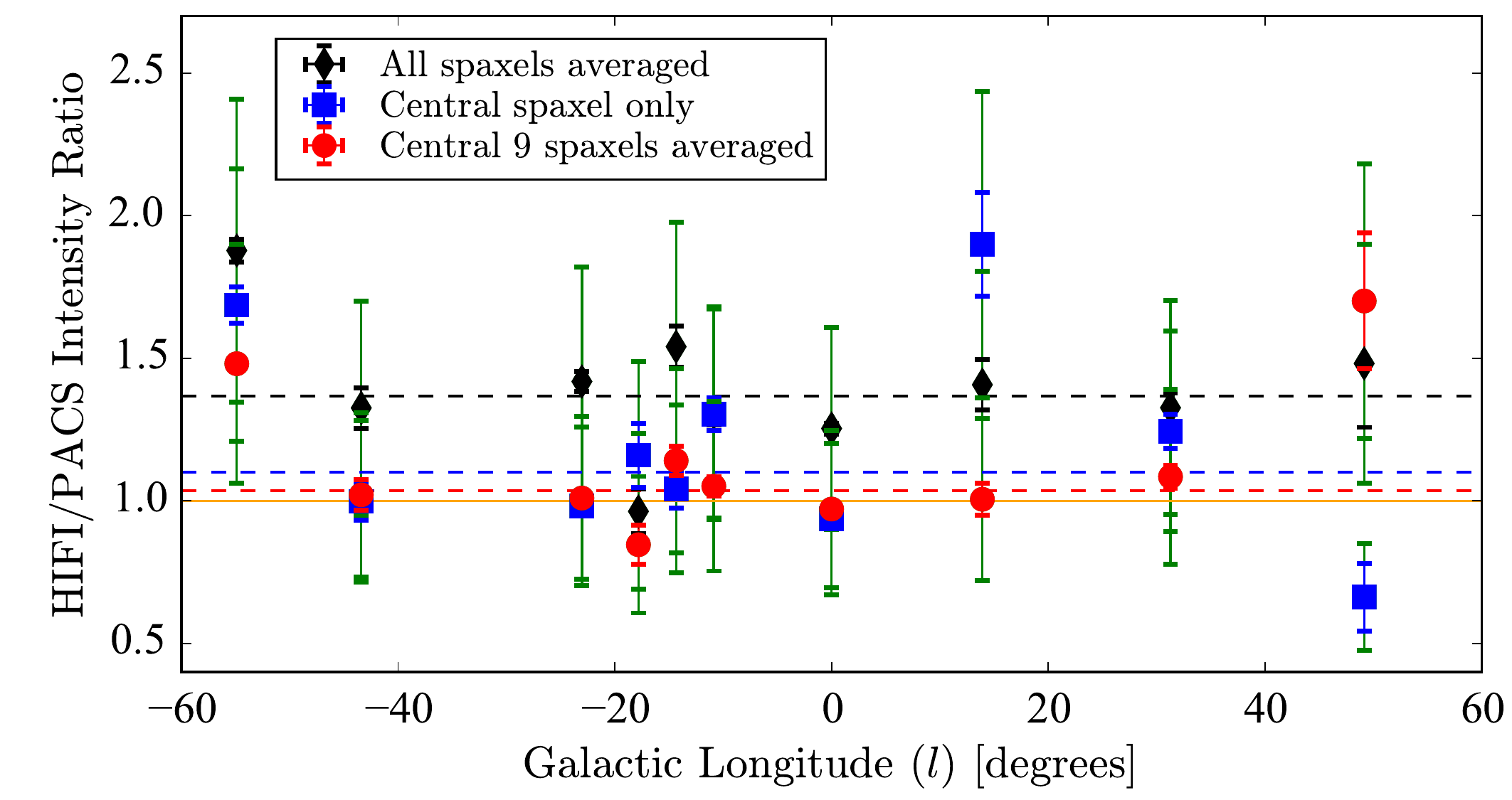}
    \caption{Comparison of PACS and HIFI 205 $\mu$m observations from the same pointing directions used for cross--calibration of the two instruments. The intensities obtained for three different configurations of PACS spaxels with increasing beam size are compared as discussed in the text. Blue, red, and black error bars are the \rms\ statistical uncertainties, and the green error bars indicate the additional 20\% calibration uncertainty added to the PACS data. The dashed lines are median values of the comparison of each configuration. The solid line denotes equal intensities for the observations with the two instruments.}
    \label{fig:PACSHIFIcomparison}
\end{figure}

After both PACS and HIFI observations are pipelined via HIPE, they are then converted to GILDAS-\verb1CLASS1\footnote{{http://www.iram.fr/IRAMFR/GILDAS/}} format for further reduction and analysis. In order to verify the calibration and reduction of both HIFI and PACS data, we have compared the intensities of the 205~\um\ line observed with the two instruments along the same line of sight. To do this, the HIFI integrated intensities were converted to intensities (\Wmtwosr) in order to allow comparison with the spectrally unresolved PACS data. The results are shown in Fig.~\ref{fig:PACSHIFIcomparison}.

The comparison was carried out for three configurations due to the different beam sizes of the two instruments. In the first, only the central PACS spaxel (which has a somewhat smaller beam size than does HIFI) was taken for comparison.  In the second configuration, 9 central PACS spaxels were averaged together resulting in a larger beam size, and in the third configuration, all 25 spaxels were averaged together.   In the first configuration, the mean HIFI/PACS intensity ratio is 1.19. For the 9 central spaxels, the mean ratio is 1.13, and for all panels, the mean HIFI/PACS ratio is 1.39 (the median values are 1.10, 1.04, and 1.39, respectively). Given the plausible variations in the source intensity over the angular size of the PACS footprint (47$\arcsec$ corresponding to $\simeq$ 1 pc at a distance of 4 kpc), this relatively small variation shows that the calibration and data reduction procedures of both instruments are consistent. We do not feel sufficiently confident in the superiority of the calibration of either instrument to change its calibration, so that we have not made any calibration adjustment.  The calibration uncertainty is much greater for the 205~\um\ data than for the 122~\um\ data. We have added a calibration uncertainty of 20\% for the PACS intensities, which is denoted by the green error bars in Fig.~\ref{fig:PACSHIFIcomparison}.

\subsection{Other data}

Our \NII\ observations were complemented by \CII\ data obtained from the {\it Herschel}  \gotclong\ (GOT C+) open time key project \citep{Langer10}. 
The data, reduction, and analysis are presented in detail by \citet{Pineda13}.

\section{RESULTS}
\label{sec:results}
\subsection{Detection Statistics and Intensities}
 
\begin{figure}[!t]
     \centering
      \includegraphics[scale=0.6]{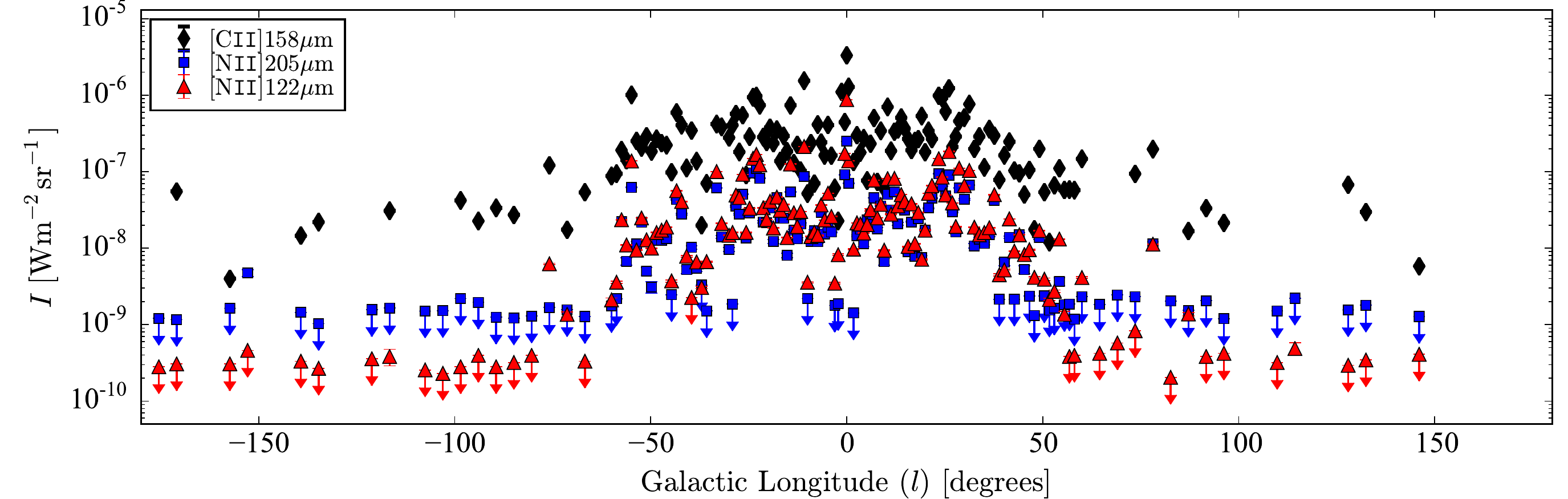}
      \caption{\small Observed intensities of \CII\ 158~\um\, and \NII\ 122~\um\ and 205~\um\ are shown as a function of Galactic longitude.}
      \label{fig:PositionFlux}
 \end{figure}

\begin{figure}[!t]
    \centering
    \includegraphics[scale=0.7]{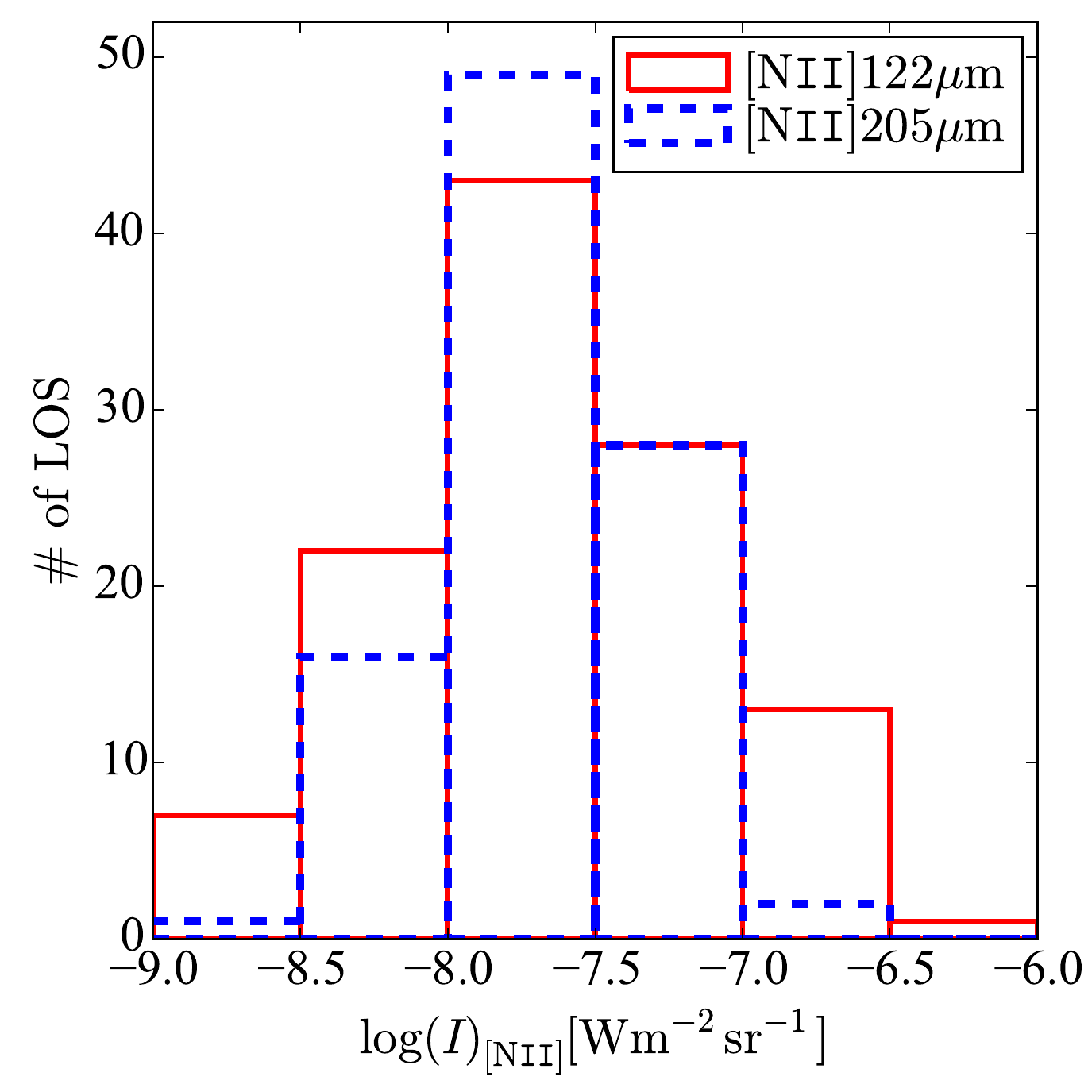}
    \caption{\small Distribution of intensities obtained from the detected LoS positions is shown by histograms of \NII\ 122~\um\ (blue) and 205~\um\ (red).}
    \label{fig:HistNII}
\end{figure}

Towards the inner Galaxy, both \NII\ lines were generally detected with high statistical significance.    However,  toward the outer Galaxy, we had mostly non-detections. Using only the central spaxel of PACS, we obtained 94 detections in the 122~\um\ observations and 59 detections in the 205~\um\ observations. In order to improve the signal to noise ratio and thus the detection statistics, we averaged together the 25 spaxels observed in each pointing.  This increased the number of detections to 116 positions at 122~\um, and 96 positions at 205~\um. Figure~\ref{fig:PositionFlux} shows the distribution of observed \NII\ and \CII\ line intensities as a function of Galactic longitude.  In Fig.~\ref{fig:HistNII}, we show a histogram of intensities in \Wmtwosr, obtained from the detected LoS positions. The median intensity for both \NII\ 122~\um\ and for \NII\ 205~\um\ is $\sim$2$\times$10$^{-8}$ \Wmtwosr.

The highest intensity of the 205~\um\ \NII\ line is measured towards the Galactic Center at G000.00+0.0, for which $I(205)$ = 2.5$\times$10$^{-7}$~\Wmtwosr. In this same direction, \citet{Bennett94}, using the COBE FIRAS instrument with a 7$\degr$ beam size measured an intensity 1.2$\times$10$^{-8}$ \Wmtwosr.  \citet{Bennett94} found a component of the \NII\ emission to be broadly extended in latitude, but this appears to produce only a small fraction of the total emission towards the Galactic center. Thus the large COBE beam dilutes the strong emission from the center of the Milky Way, making our measured intensity consistent with theirs.  The intensity of the strong \NII\ 122~\um\ emission from the direction of the Galactic Center is also in good agreement with that measured using {\it ISO} LWS by \citet{Baluteau03}, using the information provided by \citet{Lloyd03} and assuming a uniform, extended source.

\subsection{Variation Among Spaxels in a Pointing}
\label{variation}

\begin{figure}[!t]
    \centering
	\includegraphics[scale=0.5]{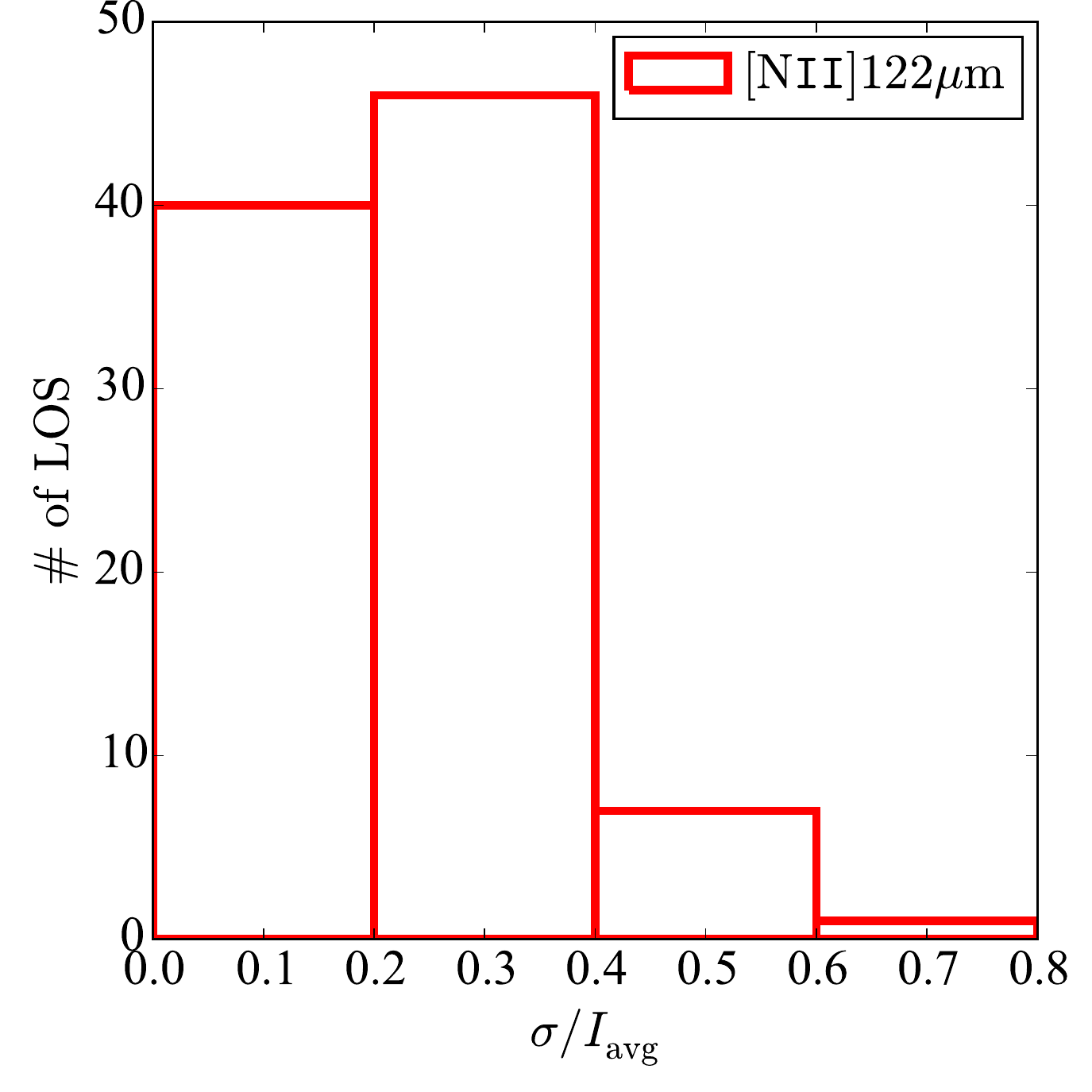}
    \caption{Histogram of   $\sigma/I_{\rm avg}$ at 122~\um\ wavelength determined from the 25 spaxels in each PACS pointing.}
    \label{fig:HistDist}
\end{figure}

The PACS spaxels are almost independent measurements of the \NII\ intensities towards 25 closely--packed directions.  We can thus consider the variation among the 25 spaxels as a measure of the uniformity of the emitted intensity. We assess the variation among the spaxels by considering the lines of sight for which we have a statistically significant detection in the central spaxel in the 122~\um\ line.  
The uncertainty in the integrated intensity in a single pixel is obtained from the intensity at wavelengths without line emission.  
The integrated line intensity and the uncertainty averaged over the 25 pixels for a given LoS are denoted ``Intensity'' and ``Err'' in columns 5 \& 8 (for 122 \um\ and 205 \um) of  Tables 2 \& 3.  
To estimate the variation of the intensity among the pixels of each LoS, it is convenient to denote the average intensity, $I_{\rm avg}$, and its standard deviation, denoted $\sigma$; the resulting values are given in column 6 of Tables 2 \& 3.  
Figure~\ref{fig:HistDist} shows a histogram of the fractional variation, $\sigma/I_{\rm avg}$.  

The variation is significantly greater than that expected from the statistical uncertainties alone, and thus represents actual spaxel--to-spaxel variations in the emission.  However, as seen in Fig.~\ref{fig:HistDist}, the fractional variation of the emission is generally well below unity, with a median value $<\sigma/I_{\rm avg}>$~=~0.25. Even the single pointing direction with the largest value of $\sigma/I_{\rm avg}$~=~0.78 has detections in $\simeq$ 1/3 of the spaxels.  It is thus reasonable to consider the \NII\ emission to be spatially extended and fairly uniform in each pointing direction ($\simeq$ 1 arc minute in size).  Given that the PACS footprint size is 47\arcsec\ and that 1\arcmin\ corresponds to 1.5 pc at a distance of 5 kpc, it certainly does not seem that the \NII\ emission in general is produced by compact sources.  

\begin{figure}[tb]
    \centering
    \includegraphics[scale=0.67]{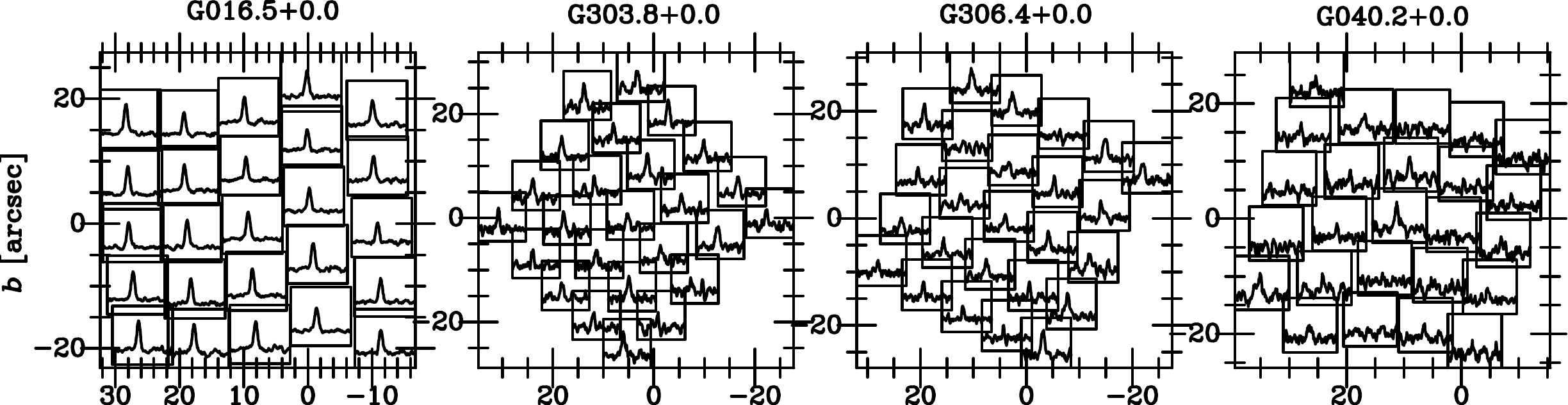}
    \includegraphics[scale=0.58]{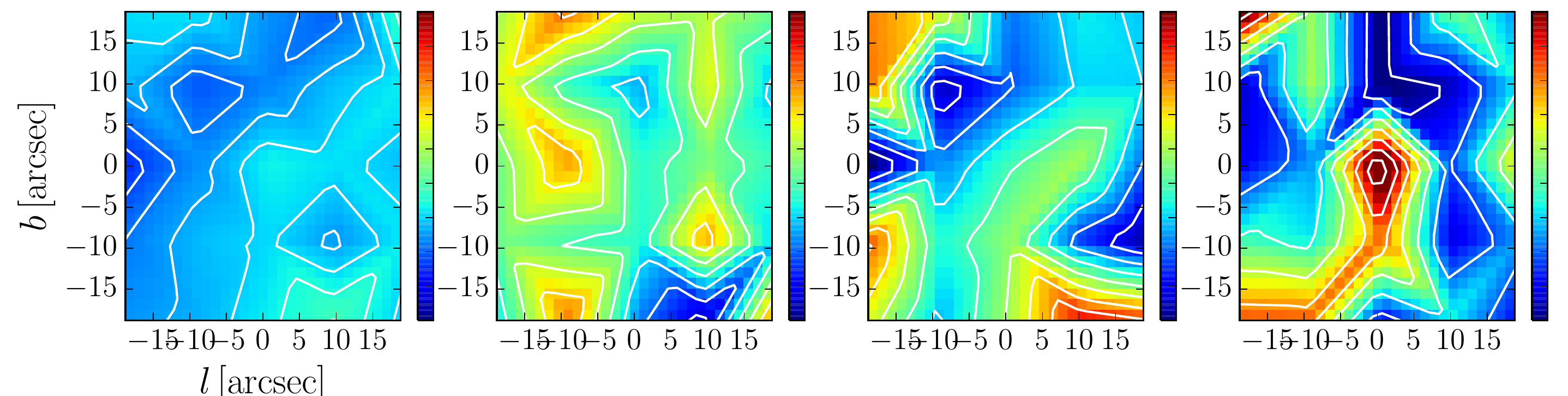}
     \caption{{\it Top:} A spectral map of 25 PACS spaxels at 122~\um\ for pointings representative of each of the segments in the histogram in Fig.~\ref{fig:HistDist}. {\it Bottom:} Corresponding contour map for each pointing. The color bar for each covers an intensity variation of one order of magnitude, with the lowest values being $I_{\rm min}$ (in units of 10$^{-8}$ \Wmtwosr) = 1.0 for G016.5+0.0, 2.0 for G303.8+0.0, 0.2 for G306.4+0.0, and 0.1 for G040.2+0.0. }
     \label{fig:PACSimages}
\end{figure}

We can gain additional appreciation of the extended nature and relative uniformity of the \NII\ emission by examining the results from selected PACS pointings.   In the top row of Fig.~\ref{fig:PACSimages}, we show a set of PACS spectra from one pointing in each of the four ranges of $\sigma/I_{\rm avg}$ shown in Fig \ref{fig:HistDist}. 
The values of $\sigma/I_{\rm avg}$ are 0.15, 0.30, 0.49, and 0.71 for G016.5+0.0, G303.8+0.0, G306.4+0.0, and G040.2+0.0, respectively.
The bottom row of Fig.~\ref{fig:PACSimages} displays contour maps of the same four pointings.  Only for G040.2+0.0 is there clear evidence of small scale structure, which is likely due to a small source close to the (0,0) position.
Because the emission is extended and at least moderately smoothly distributed, we will model the emission as produced in a layer with a single set of physical conditions (temperature, density).  This model is discussed in the following section.  The  effects of multiple regions with different conditions along the line of sight are discussed in Section \ref{multiple}.

\section{EXCITATION OF \NII\ FINE STRUCTURE TRANSITIONS}
\label{sec:excitation}

In this section, we discuss modeling of the collisional excitation of the \nplus\ fine structure transitions.  While much of this material can be found in the literature, we expand the discussion in some detail to develop relationships that are useful for analyzing the \NII\  data, determining the electron density and calculating the \nplus\ column density from the {\it Herschel} data.

The three \nplus\ fine structure levels are \thPzero, \thPone, and \thPtwo, which we denote levels, 0, 1, and 2 in order of increasing energy.  The statistical weights are $g_0$ = 1, $g_1$ = 3, and $g_2$ = 5. 
Due to its having higher ionization potential than hydrogen, ionized nitrogen is expected to be found only in highly--ionized regions (e.g., WIM, \HII\ regions), where excitation of the fine structure levels will be primarily by electron collisions.  This process has been studied in some detail by \cite{lennon94} and subsequently by \cite{Hudson04}.  More recently, a yet more detailed calculation has been carried out by \cite{Tayal11}.  At  a temperature of 8000 K, the deexcitation rate coefficients obtained from the calculations by \cite{Tayal11} are $R_{21}$ = 2.8$\times$10$^{-8}$ \cmcsone, $R_{10}$ = 1.2$\times$10$^{-8}$ \cmcsone, and $R_{20}$ = 3.3$\times$10$^{-9}$ \cmcsone.
The $^3$P$_2$ -- $^3$P$_1$ ($R_{21}$) rate coefficient in the newest calculation is approximately 33\% larger, the $^3$P$_1$ -- $^3$P$_0$ ($R_{10}$) rate coefficient is approximately 8\% smaller, and the $^3$P$_2$ -- $^3$P$_0$ ($R_{20}$) rate coefficient is aproximaely 34\% smaller than those of \citet{Hudson04}.

The result is that the electron densities derived using the most recent rate coefficients are between 5\% larger (for electron densities $\leq$ 13 \cmthree) and 20\% smaller (for electron densities $\geq$ 100 \cmthree) using the \citet{Tayal11} collision rate coefficients compared to those obtained using those of \citet{Hudson04}.  The effect on the details of the excitation are somewhat more complex as discussed further below.

\subsection{Rate Equations and Level Populations}

The populations of the three \nplus\ fine structure levels are determined by a set of three rate equations plus a constraint equation on the total density of this ion ($n({\rm N}^+)$), as laid out in early treatment by \citet{Simpson75} and others.  We consider here only spontaneous emission and collisions, as stimulated emission by internal or external sources is generally unimportant.  The optical depth of a continuum source is modest at these wavelengths, and the small solid angle subtended by a plausible external source is small, so that the impact on the radiation density at the frequencies of the \NII\ transitions can be neglected.  We also consider the lines to be optically thin; this assumption will be discussed further below, but both theoretically and observationally seems to be justified.

We denote the three fine structure levels by their total angular momentum ($J$), their densities (\cmthree) of the three fine structure levels as $n_0$, $n_1$, and $n_2$.
For electron density \ne, the collision rates from level $i$ to level $j$, $C_{ij}$, are given by
\begin{equation}
C_{ij} = R_{ij}n(e) \lc
\end{equation}
where $R_{ij}$ is the collision rate coefficient (\cmcsone) and $n(e)$ is the electron density (\cmthree).

The rate and constraint equations can be written as
\begin{equation}\label{rates}
\begin{split}
-(A_{21} + C_{21} + C_{20})n_2 + C_{12}n_1 + C_{02}n_0 &= 0\\
(A_{21} + C_{21})n_2  - (A_{10} + C_{10} + C_{12})n_1 + C_{01}n_0 &= 0\\
C_{20}n_2 + (A_{10} + C_{10})n_1 - (C_{01} + C_{02})n_0 &= 0\\
n_2 +n_1 + n_0 &= n({\rm N}^+)
\end{split}
\end{equation}

There are many ways to describe the solution to the above set of equations, but since for other purposes the ratio of the populations is needed, we first show the two ratios of the populations of adjacent levels (connected by radiative transitions), which are
\beq
\label{R21}
R(2/1) = \frac{n_2}{n_1} = \frac{C_{12}(C_{01} + C_{02}) + C_{02}(A_{10} + C_{10})} {(A_{21} + C_{21} + C_{20})(C_{01} + C_{02}) - C_{20}C_{02}} \lc
\eeq
and
\beq
\label{R10}
R(1/0) = \frac{n_1}{n_0} = \frac{(A_{21} + C_{21} + C_{20})(C_{01} + C_{02}) - C_{20}C_{02}} {(A_{21} + C_{21} + C_{20})(A_{10} + C_{10}) + C_{20}C_{12}} \lp
\eeq
It is evident that by multiplying the two preceding expressions together, one can obtain the ratio of $n_2$ to $n_0$.  

In the low--density limit $C\ll A$, Equation~\ref{R21} becomes
\beq
\label{Rlowdens}
R(2/1)|_{C\ll A} = \frac{C_{02}A_{10}}{(C_{01} + C_{02})A_{21�}} ~,
\eeq
so that the ratio of the two upper level populations approaches a constant value of 0.11 at low densities.

\begin{figure}
\begin{center}
\includegraphics[width = 0.8\textwidth]{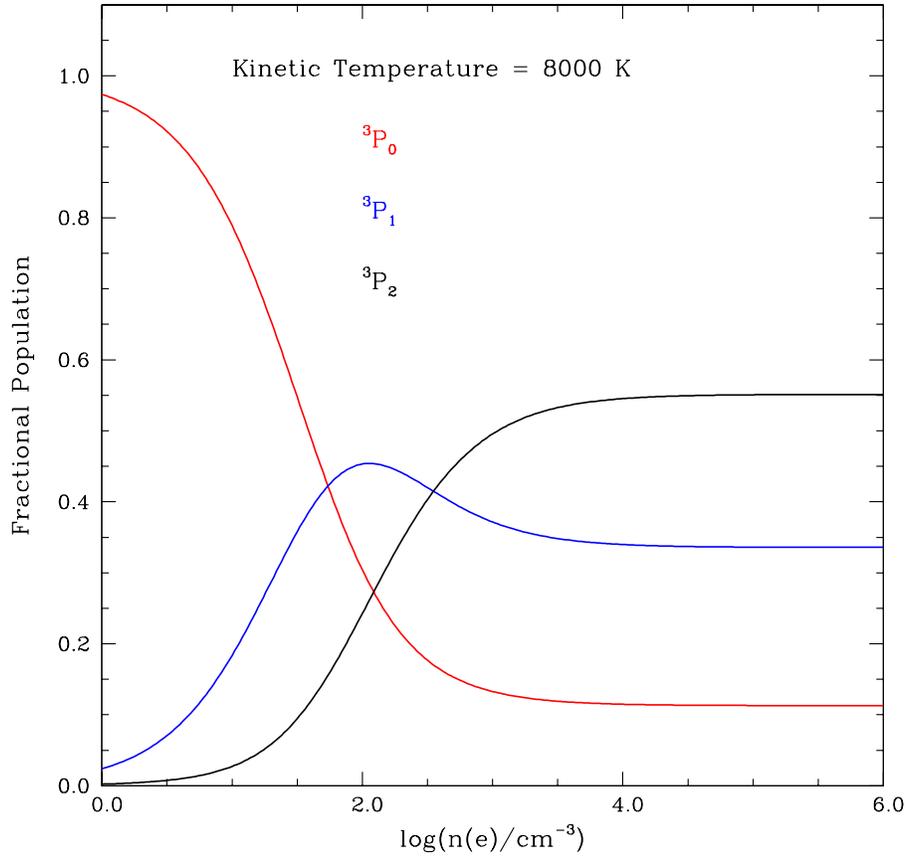}
\caption{\label{fig:fractional_pops} Fractional populations of three \nplus\ fine structure levels for kinetic temperature of 8000 K, as a function of electron density.}
\end{center}
\end{figure}

From Equations~\ref{R21} and \ref{R10}, the fractional population in each of the three levels can be obtained from
\begin{equation}\label{fractions}
\begin{split}
\frac{n_2}{n_{\rm N^+}} &= (1 + R(2/1)^{-1} + R(2/1)^{-1}R(1/0)^{-1})^{-1}\\
\frac{n_1}{n_{\rm N^+}} &= (1 + R(2/1) + R(1/0)^{-1})^{-1}\\
\frac{n_0}{n_{\rm N^+}} &= (1 + R(1/0) + R(2/1)R(1/0))^{-1}
\end{split}
\end{equation}
The variation of the three fractional populations with electron density is shown in Fig.~\ref{fig:fractional_pops}, for a kinetic temperature of 8000 K using the \cite{Tayal11} collisional rate coefficients.
In the low--density limit essentially all of the population is in the lowest level, and the fractional populations of the two excited levels are
\begin{equation}\label{lowdens_fractions}
\begin{split}
\frac{n_2}{n_{\rm N^+}} &= \frac{C_{02}}{A_{21}}\\
\frac{n_1}{n_{\rm N^+}} &= \frac{C_{01} + C_{02}}{A_{10}}\\
\end{split}
\end{equation}
This solution necessarily gives the same limiting value for the level population ratio as given in Equation~\ref{Rlowdens}.  It also makes clear that the population and hence the intensity of both fine structure transitions is linearly proportional to the electron density in the low--density limit.

\subsection{Excitation Temperature and Critical Density}

\begin{figure}
\begin{center}
\includegraphics[width = 0.8\textwidth]{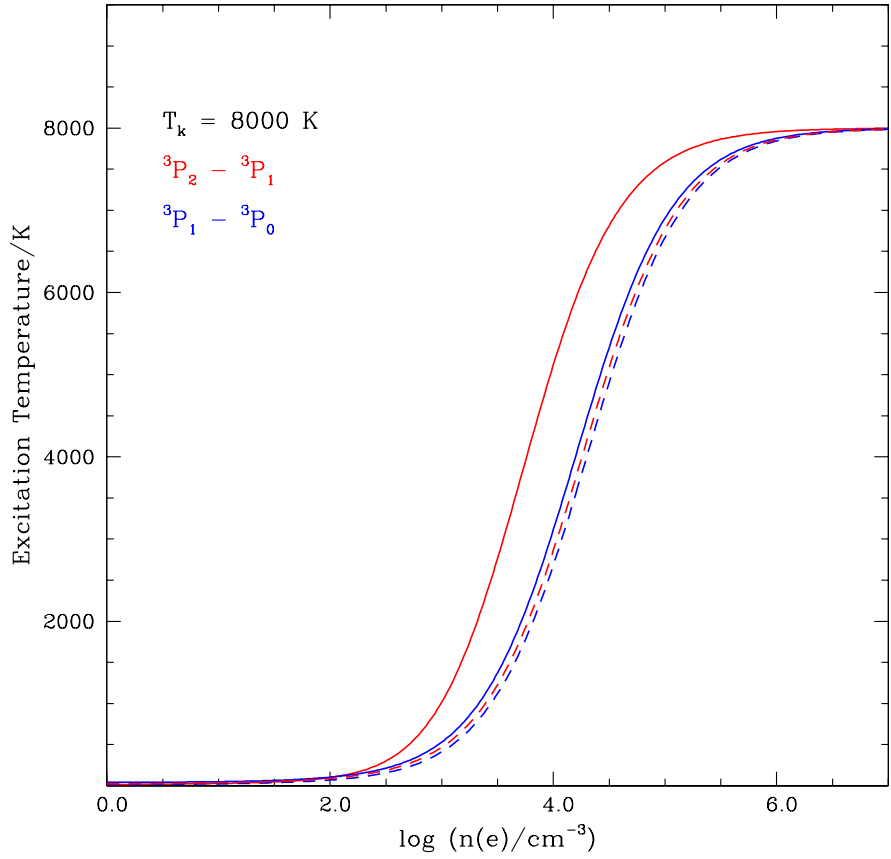}
\caption{\label{fig:tex} Excitation temperatures of \nplus\ fine structure transitions as a function of electron density, for kinetic temperature of 8000 K.  The solid curves are for the \citet{Tayal11} collision rate coefficients while the dashed curves have the same coefficients, but with the \thPtwo\ -- \thPzero\ rate set equal to zero.}
\end{center}
\end{figure}

The excitation temperature of a transition is a measure of the relative population of its upper ($j$) and lower ($i$) levels, defined through
\beq
\label{tex}
R(j/i) = \frac{n_j}{n_i} = \frac{g_j}{g_i}exp[-(E_j - E_i)/kT_{{\rm ex}~ij}] \lc
 \eeq
 where the $g$'s are the statistical weights and $E$'s the energies of the levels.  Defining $T^*_{ij}$ = $(E_j - E_i)/k$ , we see that
 \beq
 T_{{\rm ex}~ij} = \frac{-T^*_{ij}}{ln[\frac{g_i}{g_j}R(j/i)]} \lp
 \label{tex2}
 \eeq
The \cite{Tayal11} rate coefficients, as well as the earlier calculations, indicate that all three levels are coupled by collisions.  We can use equations~\ref{R21} and \ref{R10} to determine the excitation temperatures of the two \nplus\ fine structure lines, and these are plotted in Fig.~\ref{fig:tex}. It is evident in the figure that the excitation temperatures are very low for low electron densities  and that both become equal to the kinetic temperature (LTE) at sufficiently high electron densities ($\ge$ 10$^5$ cm$^{-3}$).  

We can gain some insight into this behavior by first considering collisions that connect only adjacent levels.  In this case, the excitation temperatures are decoupled, and for each transition we can write (for optically thin lines and neglecting any background)

\beq
T_{{\rm ex}~ij} = \frac{T^*_{ij}} {\frac{T^*_{ij}} {T_{\rm k}} + ln(1 + \frac{A_{ji}}{C_{ji}})} \lc
\label{tex3}
\eeq
where $T_{\rm k}$ is the kinetic temperature.

The critical density is unambiguously defined for a two level system as the density at which the collisional deexcitation rate is equal to the spontaneous decay rate.  With this definition, which gives a measure of the density at which the transition is significantly excited, the critical density can be written
\beq
n_{{\rm e~cr}~ji} = \frac{A_{ji}}{R_{ji}}\lp
\eeq
This definition can readily be extended to include the effects of radiative trapping through an ``effective decay rate'', typically written as $A_{\rm eff} = \beta A_{ji}$, where $\beta$ is the photon escape probability.

The excitation temperature in equation \ref{tex3} can be written in terms of the critical density as
\beq
T_{{\rm ex}~ij} = \frac{T^*_{ij}} {\frac{T^*_{ij}} {T_{\rm k}} + ln(1 + \frac{n_{\rm e~cr~ji}}{n_{\rm e}})} 
\eeq

The critical density has a more physical interpretation, which becomes evident by considering a two level or a decoupled mutilevel system as described by equation \ref{R21} or \ref{R10} with only adjacent levels coupled by collisions.  Rather than involving the kinetic temperature, we write the ratio the populations as obtained from the rate equation as
\beq
\frac{n_j}{n_i} = \frac{C_{ij}}{C_{ji} + A_{ji}} = \frac{g_j}{g_i}\frac{e^{-T^*_{ij}/T_{\rm k}}}{1 + A_{ji}/C_{ji}}
= \frac{g_j}{g_i}\frac{e^{-T^*_{ij}/T_{\rm k}}} { 1 + n_{{\rm e~cr}~ji}/n_{\rm e} }\lp
\eeq
We see that the ratio of the population of the upper level to that of the lower level in LTE ($C_{ji} \gg A_{ji}$ or $n_{\rm e} \gg n_{\rm e~ cr}$) is just $(g_j/g_i)e^{-T^*_{ij}/T_{\rm k}}$ and that the population ratio for $C_{ji}$ = $A_{ji}$ or equivalently $n_{\rm e}$ = $n_{\rm e~cr}$ is just one half of the LTE value.  Thus, the critical density is that which results in a collision rate that makes the ratio of upper to lower level population one half of its value in LTE at the kinetic temperature.  This definition is very meaningful inasmuch as it defines an appropriate fraction of the maximum possible level population for a given total column density of the species in question.  It also defines a useful measure of the optical depth relative to its LTE value, which is clearly of importance when analyzing absorption lines.  These attributes are independent of the issue of the ``detectability'' and the ``effective density'' of a given species, as discussed by \cite{evans99} and \cite{shirley15}.

The situation for multilevel systems can be more complex, because in general collisions couple the populations not just of adjacent levels, but of many levels with arbitrary separation.  This situation is illustrated by the rotational levels of the CO molecule, for which collisions with $\Delta J$ exceeding 5 can be significant \citep{Yang10}.

For \nplus\ we have only three levels, but they are in reality all collisionally coupled as discussed above.  The population ratios can be written in a form that emphasizes the similarity with a two level system as
\beq
\label{ind21}
R(2/1) = \frac{n_2}{n_1} = \frac{C_{12} + (A_{10} + C_{10})\frac{C_{02}}{C_{01} + C_{02}}} { A_{21} + C_{21} + C_{20}\frac{C_{01}}{C_{01} + C_{02}}} \lc
\eeq
and
\beq
\label{ind10}
R(1/0) = \frac{n_1}{n_0} = \frac{C_{01} + C_{02}\frac{A_{21} + C_{21}} {A_{21} + C_{21} + C_{20}} }
  {A_{10} + C_{10}  + C_{12}\frac{C_{20}} {A_{21} + C_{21} + C_{20}} } \lp
\eeq
For each ratio, the first term in the numerator is the direct (collisional) excitation rate, and the first two terms in the denominator are the collisional and the radiative deexcitation rates.  The remaining terms are the indirect paths for population transfer.  For example, in $R(2/1)$, $C_{10} + A_{10}$ is the total rate from level 1 to level 0, while $C_{02}/(C_{01}+ C_{02})$ is the fraction of that rate resulting in transfer of population to level 2.  
From Equations~\ref{ind21} and \ref{ind10}, it is evident that each population ratio does not depend exclusively on the ratio of its collisional and radiative rates, unless the terms $C_{20}$ and $C_{02}$ are zero.  Thus, it is not clear how the critical density should be defined. 

One way to get a feeling for the effect of the indirect terms is to consider that $C_{10}$ = $A_{10}$ in evaluating $R(2/1)$ and that $C_{21}$ = $A_{21}$ in evaluating $R(1/0)$.  With the known values for the upwards and downwards rates, we find that including the indirect routes results in an {\bf effective increase} in the collision rate of $\simeq$ 15\% when evaluating $R(2/1)$ and of $\simeq$ 30\% for $R(1/0)$, although these results are only approximate.  

The critical densities for the two \nplus\ fine structure transitions are given in Table~\ref{ncr}, for the two different collisional selection rules.  We confirm that for $|\Delta J|$ = 1 only collisions, that the value of the critical density found using the definition of population ratio equal to half its LTE value agrees with that found by dividing the spontaneous decay rate by the collisional deexciation rate coefficient.  The reduction in the critical density produced by the complete rates compared to the $|\Delta J|$ = 1 only rates is modest for the higher frequency transition, but substantial for the lower frequency transition.  This behavior is the same for the \citet{Hudson04} and \citet{Tayal11} collisional rates, although the magnitudes of the critical densities reflect the changes in the rate coefficients discussed above.
\begin{deluxetable}{ccc}
\renewcommand\thetable{5} 

\normalsize
\tablecolumns{3}
\tablewidth{0pt}
\tablecaption{\label{ncr}  Critical Densities for Excitation of \nplus\ Fine Structure Transitions by Electrons\tablenotemark{a} }

\tablehead{\colhead{Transition} & \colhead{$n_{\rm cr}$} & \colhead{$n_{\rm cr}$} \\
                                & \colhead {$|\Delta J|$ = 1} & \colhead{All $\Delta J$}}

\startdata
$^3$P$_2$ -- $^3$P$_1$ (2 -- 1) &264 &220\\
$^3$P$_1$ -- $^3$P$_0$ (1 -- 0) &175 &100\\
\enddata

\tablenotetext{a} {In units of cm$^{-3}$. Collision rates from \cite{Tayal11} for a kinetic temperature of 8000 K}
\end{deluxetable}

\subsection{Connection with Observations}
\label{conn_obs}
\begin{figure}
\begin{center}
\includegraphics[width = 0.8\textwidth]{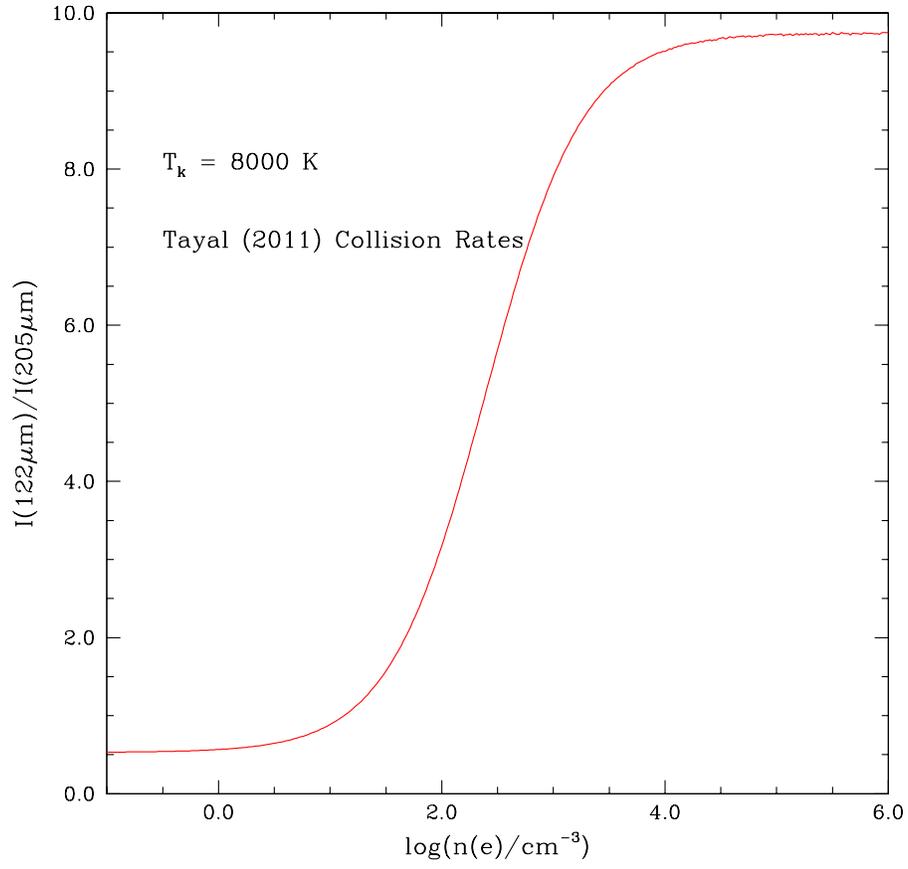}
\caption{\label{fig:R21} Ratio of intensity of the \NII\ 122~$\mu$m line to that of the 205~$\mu$m line for optically thin emission at 8000 K, as a function of electron density.}
\end{center}
\end{figure}

The brightness (or specific intensity) integrated over frequency for an optically thin spectral line of frequency $f_{ul}$, spontaneous decay rate $A_{ul}$, and upper level column density $N_u$ is given by 
\beq
I = \frac{A_{ul}hf_{ul}N_u}{4\pi} \lp
\eeq
Consequently, the ratio of the intensities of the two fine structure lines of \nplus\ can be written
\beq
\label{I_ratio1}
\frac{I_{21}}{I_{10}} = \frac{A_{21}f_{21}N_2} {A_{10}f_{10}N_1} \lc
\eeq
and assuming that the ratio of column densities is equal to the ratio of the volume densities, we obtain
\beq
\label{I_ratio2}
\frac{I_{122}}{I_{205}} = \frac{I_{21}}{I_{10}} = \frac{A_{21}f_{21}} {A_{10}f_{10}}R(2/1) \lp
\eeq
This equation gives for the \nplus\ fine structure lines
\beq
\label{I_ratio3}
\frac{I_{122}}{I_{205}} = 6.05 \, R(2/1) \lp
\eeq
We now have a straightforward method of determining the electron density from observation of the ratio of the two \nplus\ fine structure transitions, as discussed by \cite{Oberst06, Oberst11}.  Figure~\ref{fig:R21} illustrates the dependence of the observed intensity ratio on the electron density for a kinetic temperature of 8000 K.  Due to the very weak dependence of the collision rate coefficients on temperature, the result is essentially independent of the temperature, given that the emission is produced in a region in which hydrogen is largely ionized.  

Due to the asymptotic low density behavior of $R(2/1)$ given in Equation~\ref{Rlowdens}, the ratio of the two line intensities similarly approaches a constant value, equal to 0.65 for low electron densities.  At high densities, we achieve LTE at the kinetic temperature, and from Equation~\ref{tex}, $R(2/1)$ = 1.64, and $I_{122}/I_{205}$ = 9.74.  From an observational point of view, any ratio of two lines is a useful densitometer only over a restricted range of densities, which for \NII\ $I_{122}/I_{205}$ is $10^1 \leq n(e) \leq 10^3$ \cmthree.   However, if we include sensitive upper limits for non--detections, we can draw conclusions about conditions spanning a wider range of electron densities. For example, if the electron density were 1 \cmthree\ or less, the value of $I_{122}/I_{205}$ would be so low that the \NII\ 122~\um\ line would be undetectable even if the column density of \nplus\ were large enough that the 205 \um\ line would be detectable.  
%

\subsection{Analytic Solution for the Electron Density}
\label{analytic}
From Equation~\ref{R21} (assuming uniform conditions along the line of sight), we can write the  volume density ratio of \nplus\ in level 2 to level 1 in terms of the observed ratio of the transition intensities. Taking the reciprocal of Equation~\ref{I_ratio3} yields $R(2/1) = 0.165 I_{122}/I_{205}$. 
Expressing the collision rates in terms of the collision rate coefficients and the electron density, $R_{ij} = C_{ij}n(e)$, we can invert the result to solve for the electron density in terms of $I_{122}/I_{205}$. With the definitions
\beq
\begin{split}
a &= R_{12}R_{01} + R_{12}R_{02} + R_{02}R_{10} \\
b &= R_{02}A_{10}\\
c &= R_{21}R_{01} + R_{21}R_{02} + R_{20}R_{01}\\
{\rm and}\\
d &= (R_{02} + R_{01})A_{21}~,
\end{split}
\eeq
we obtain an explicit expression for the electron density $n(e)$ in terms of the observed intensity ratio $I_{122}/I_{205}$
\beq
n(e) = \frac{d}{c}~\frac{X - R_{\rm min} } {R_{\rm max} - X} \lc
\eeq
where
\beq
\begin{split}
R_{\rm min}  &= b/d\\
R_{\rm max} &= a/c\\
{\rm and}\\
X & = R(2/1) = 0.165\frac{I_{122}}{I_{205}}.
\end{split}
\eeq

\subsection{Optical Depth of the \NII\ Transitions}
\label{tau}

\begin{figure}
\begin{center}
\includegraphics[width = 0.8\textwidth]{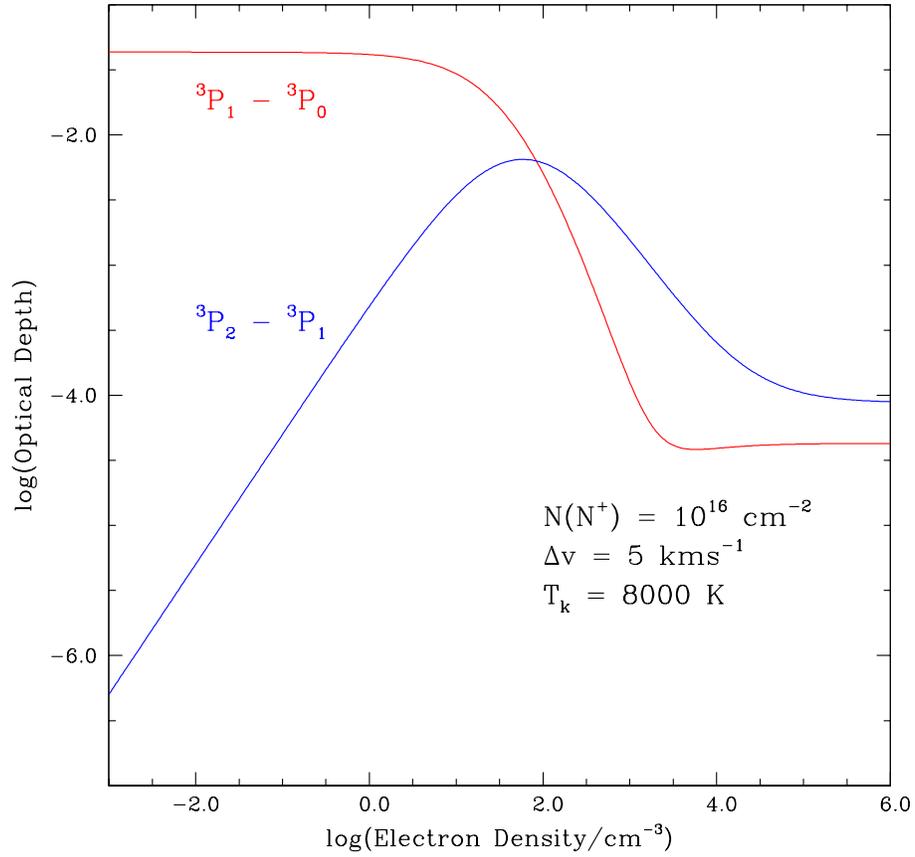}
\caption{\label{NII_tau} Optical depth of the two \NII\ fine structure transitions as a function of electron density.  The \nplus\ column density is 10$^{16}$ \cmtwo, the line width is 5 \kms, and the kinetic temperature is 8000 K.}
\end{center}
\end{figure}

The two fine structure transitions of \nplus\ behave very differently as a function of excitation, especially given that there is no significant background radiation field to populate the higher levels.  We start (for zero excitation) with all of the population in the ground state.  Consequently, the optical depth of the \thPone\ -- \thPzero\ transition is maximum.  As the excitation rate increases, the population of the $^3$P$_1$ level becomes significant, and the optical depth of the lower transition drops dramatically while that of the \thPtwo\ -- \thPone\ transition rises, reaches a maximum for $n(e)$ $\simeq$ 100 \cmthree, and then drops by almost a factor of 100 when the excitation rate is sufficient for LTE to be reached ($n(e)$ $\geq$ 10$^4$ \cmthree).  This behavior is shown in Fig.~\ref{NII_tau} for a kinetic temperature of 8000 K.  $\tau$($^3$P$_1$ -- $^3$P$_0$) drops by a factor of $\simeq$1000 from zero excitation to LTE (the exact value of the kinetic temperature is unimportant providing that it is $\gg$ $hf_{ul}/k$).  In LTE, $\tau$($^3$P$_2$ -- $^3$P$_1$) is approximately a factor 2 greater than $\tau$($^3$P$_1$ -- $^3$P$_0$).  

The maximum optical depth thus occurs for very low excitation, and it is the lower fine structure line that will first become optically thick under these conditions if the column density of \nplus\ is sufficiently large.  For the column density and line width used in the example here, $\tau$($^3$P$_1$ -- $^3$P$_0$) = 0.043 in the limit of low excitation.  From this value we derive that $\tau_{\rm max}$($^3$P$_1$ -- $^3$P$_0$) = 2.2$\times$10$^{-17}N$(\nplus)[cm$^{-2}$]/$\delta v$[\kms].  For a representative line width of 5~\kms, the maximum \nplus\ column density guaranteeing optical depth less than 0.5 for all excitation conditions is $\simeq$10$^{17}$ \cmtwo.  In a region with electron density $\geq$ 100 \cmthree, a column density as much as 10$^{18}$ \cmtwo\ can still result in both \nplus\ fine structure transitions being optically thin.  

Since the \nplus\ column densities derived from our data are virtually all less than 10$^{17}$ \cmtwo\ and derived electron densities are relatively large, it does not appear that corrections for finite optical depth are necessary.  For the WIM with much lower electron density, the maximum \nplus\ column density for optically thin lines would be the 10$^{17}$ \cmtwo\ limit, and lines can be somewhat optically thick \citep{Persson14}.

\subsection{Effect of Regions with Different Electron Densities Along the Line of Sight on Density Determination from \NII\ Fine Structure Transitions}
\label{multiple}

Although \nplus\ is found only in ionized regions, and thus is presumably less widely distributed than \cplus, it is still possible that regions having different electron densities and thus different excitation conditions exist along a given line of sight.  An obvious example would be a direction that included emission from the WIM with electron density 0.01 $\leq n(e) \leq 0.1$ \cmthree, together with a ``classical'' \HII\ region having 10 \cmthree\ $\leq$ \ne\ $\leq$ 1000 \cmthree.  It is thus of interest to examine the effect of such a combination on the electron density determined from observations of the \NII\ fine structure lines.  We will use these results in Section \ref{origins} in discussing the derived electron densities as well as the observed \nplus\ column densities.

The electron density is directly determined from the ratio of the intensity of the upper (122~\um, \thPtwo\ -- \thPone) transition to that of the lower (205~$\mu$m, \thPone\ -- \thPzero) transition.  The ratio is a monotonic function of $n(e)$ and rises from $\simeq$ 0.6 at low densities ($\leq$ 10 \cmthree) to $\simeq$ 10 for high electron densities ($\geq$ 1000 \cmthree), as discussed above.
The purpose of the following discussion is to examine what happens when there are regions with different electron densities present in the beam used for the observations.

\subsubsection{Model}

We construct a very simple toy model that includes two regions.  They are assumed to have the same kinetic temperature, 8000 K, as this value is characteristic of both the WIM and \HII\ regions. We assume that the emission is optially thin so that the intensity in each transition is linearly additive.  We take Region 1 to be the ``high density region'' (representative of an \HII\ region or cloud edge ionized by a massive young star) and Region 2 to be the ``low density region'' (representative of the WIM).  We fix the \nplus\ column density of Region 1, which we assume to have $n(e)$ = 10, 100, or 1000 \cmthree, and vary the column density of Region 2, which we assume to have $n(e)$ = 0.01 or 0.1 \cmthree.  

\begin{figure}[!]
\begin{center}
\includegraphics[width = 0.8\textwidth]{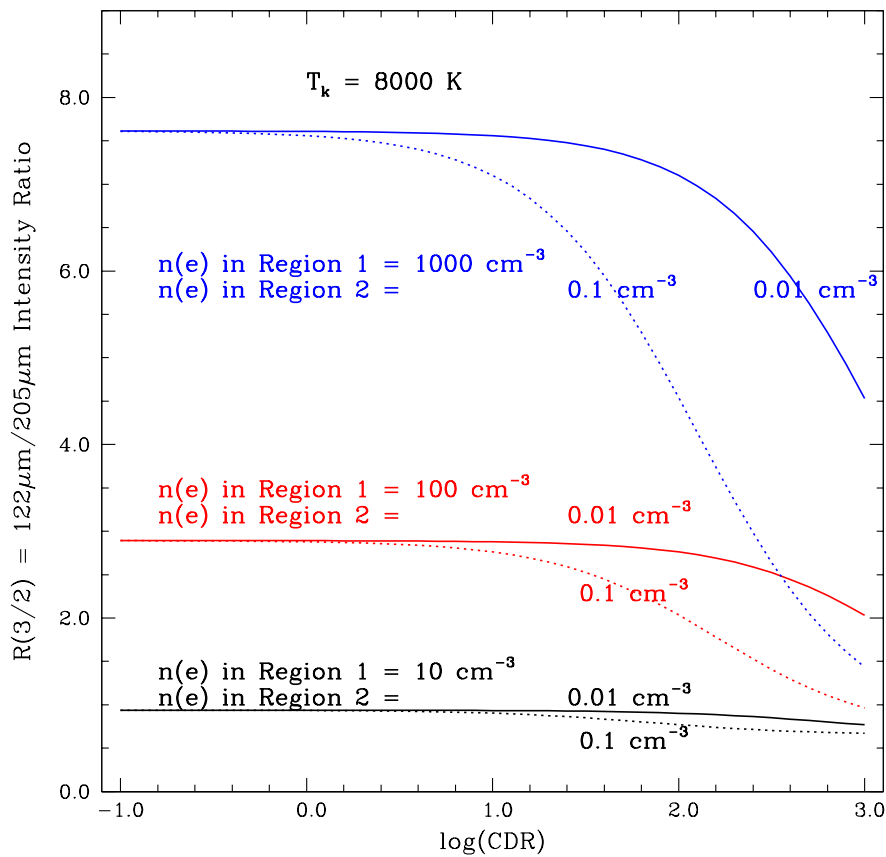}
\caption{\label{NII_comb} Ratio of intensities of the two \NII\ fine structure transitions produced by combination of two regions, Region 1 and Region 2, each optically thin and filling the telescope beam.  The kinetic temperature in both regions is 8000 K, and the electron density in each region is indicated in the figure.  The \nplus\ column density ratio in the two regions is denoted by $CDR$.  }
\end{center}
\end{figure}

The ratio of the \nplus\ column density in Region 2 to that of Region 1, the column density ratio, is denoted $CDR$ and we consider --1 $\leq$ log($CDR$) $\leq$ 3.  The column densities themselves are unimportant as long as both transitions are optically thin.  We solve for the fractional populations of each level in each of the two regions, multiply by the fraction of the total column density in each region, and calculate the intensity of each transition emerging from the combination of the two regions.  The results of the model are shown in Fig. \ref{NII_comb}.  

\subsubsection{Results of Model Calculation}
\label{model_results}

At the left of the figure, ($CDR \leq$ 1) we obtain the intensity ratio characteristic of Region~1 alone.  The perhaps initially surprising result is that as $CDR$ increases, there is almost no detectable effect on the observed ratio until $N$(Region 2)/$N$(Region 1) approaches 10 for $n(e)$ in Region 1 equal to 0.1 \cmthree, and for $CDR$ a factor of 10 larger for $n(e)$ in Region 2 equal to 0.01 \cmthree.  This behavior is a consequence of the fact that for low electron density, neither of the two excited levels has significant fractional population.  Thus, it takes a very large column density ratio to begin to affect the total column density in either the \thPtwo\ or the \thPone\ level.  For extremely large $CDR$, the observed intensity ratio will indeed approach that of Region 2 alone, but in most realistic situations the observed intensity ratio will be essentially that of Region 1.

The modest change in the observed intensity ratio is, as expected, to lower the value from that of Region 1 towards that of Region 2.  However, the effect is quite modest until we get to very large $CDR$.  As an example, consider Region 1 with $n(e)$ equal to 100 \cmthree, which results in an observed intensity ratio of 2.9.  Region 2 with a factor of 100 greater \nplus\ column density reduces the observed ratio to 2.8 if it has an electron density of 0.01 \cmthree, characteristic of the WIM.  If the electron density in Region 2 is 0.1 \cmthree, the intensity ratio is reduced somewhat more, to 2.03.  Using these observed intensity ratios and assuming a {\it single} density along the line of sight, we would derive $n(e)$ =  93 \cmthree\ with lower density Region 2 present, and $n(e)$ = 55 \cmthree\ with the higher density Region 2, both not very far below that of the high--density Region 1, 100 \cmthree.
  
 We conclude that even for a column density ratio of 100, the derived electron density differs from that of the high density region alone by less than a factor of 2.  The effect of the lower density region is to reduce the derived electron density, but the effect is clearly modest.  When we derive an electron density, we are very likely highly dominated by that of the high density region along the line of sight and the \nplus\ column density similarly will be close to that of the high density region.  This result is a more precise statement of the fact that for densities characteristic of the WIM or not much greater, the fractional population of the upper levels of the observed transitions is very small, and consequently the observed emission is dominated by a small fractional column density of higher density material.  The present results are consistent with the expectation that gas having density closer to the critical density is the most efficient at producing photons.

\section{Electron Densities and $N^+$ Column Densities Determined from \NII\ Observations}
\label{electron_dens}

\subsection{Derived Electron Densities}
\begin{figure}[!t]
    \centering
    \includegraphics[scale=0.7]{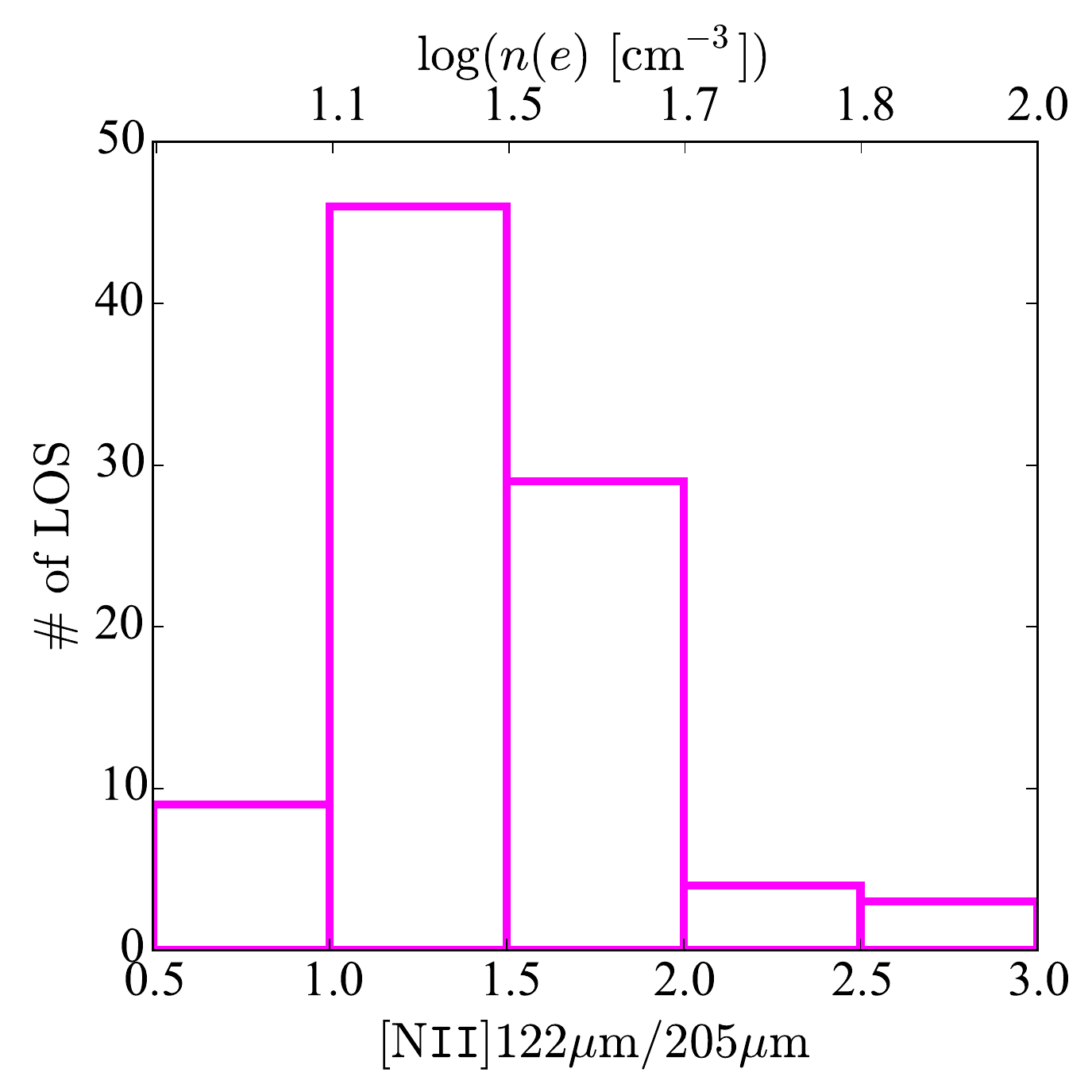}
    \caption{\small Distribution of observed \NII\ line intensity ratios. The upper x--axis gives the corresponding electron density.}
    \label{fig:HistNIIratio}
\end{figure}

For each pointing direction, we have determined the electron density from the ratio of the two \NII\ fine structure lines observed.  We have assumed that the entire emission profile is produced by a region with a single electron density.  Then, from the ratio of the 122~\um\ line to the 205~\um\ line, \ne\ can be directly determined.  The results are given in Tables 2 \& 3.
(column 9), with accompanying uncertainties.  Figure~\ref{fig:HistNIIratio} shows the distribution of the  line intensity ratio and resulting \ne.  The mean intensity ratio is 1.50.  The upper axis of Fig.~\ref{fig:HistNIIratio} is the electron density.  The mean electron density from the present data is $<n(e)>$~=~29~\cmthree.

\begin{figure}[!t]
    \centering
    \includegraphics[scale=0.7]{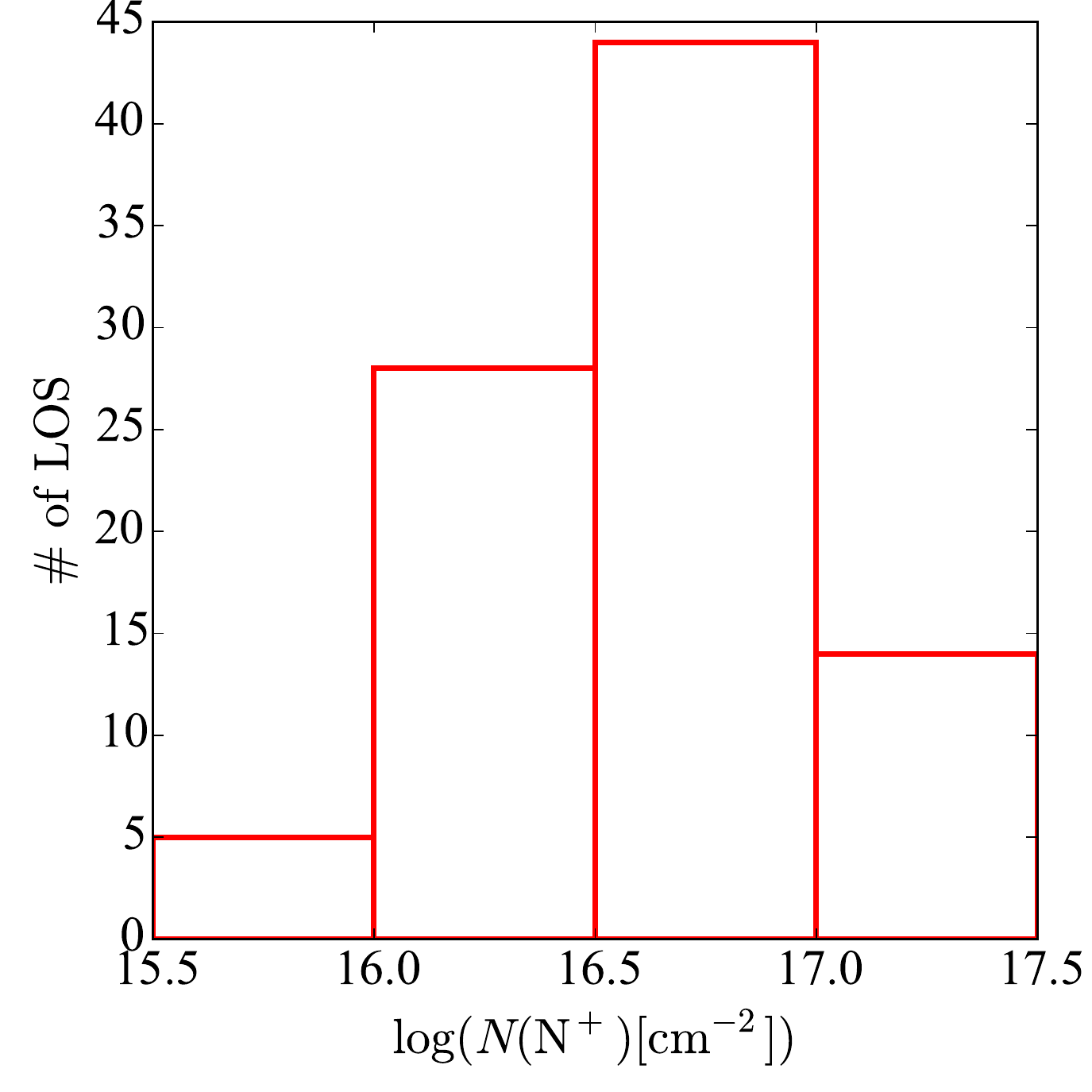}
    \caption{\small Distribution of observed N$^+$ column densities.}
    \label{fig:HistNNII}
\end{figure}

\begin{figure}[!t]
\centering
\includegraphics[scale=0.6]{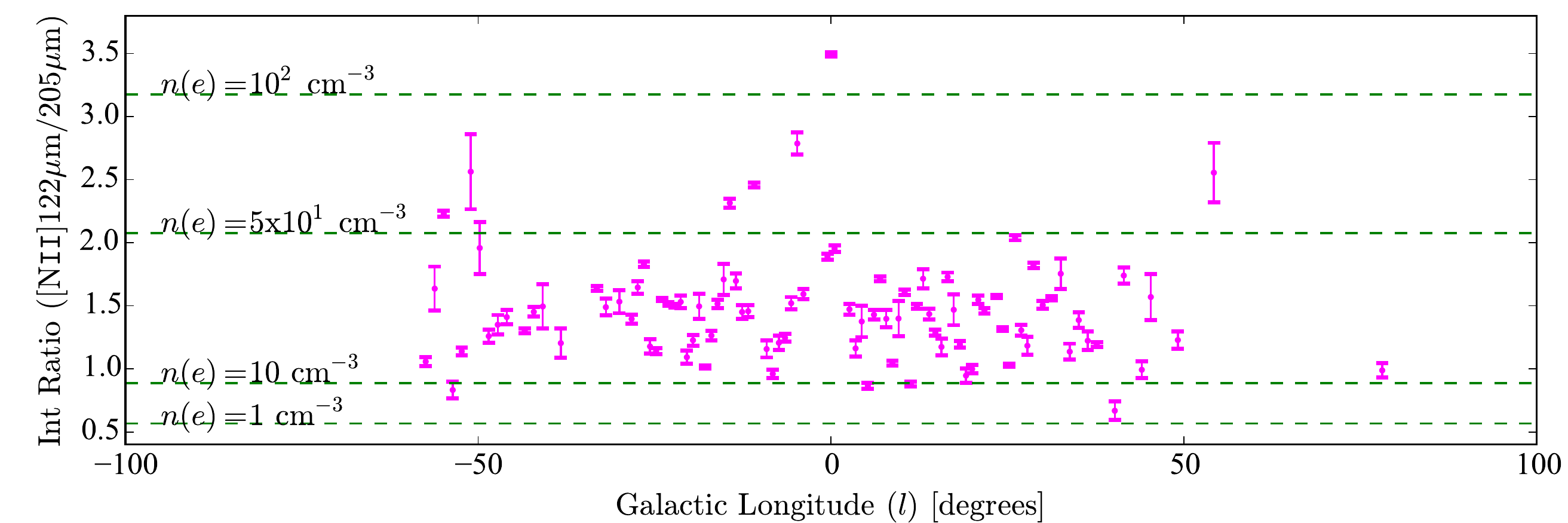}
\caption{ Observed intensity ratio of the two \NII\ fine structure lines and derived electron densities are plotted as a function of Galactic longitude.}
\label{fig:ne_vs_l}
\end{figure}

Measurement of the intensity ratio determines the electron density and the population in each fine structure levels and thus 
the total N$^+$ column density (given in Tables~2 and 3, 
column 10).  We show the distribution of column densities in Fig. \ref{fig:HistNNII}.  The mean N$^+$ column density is 4.7$\times$10$^{16}$ \cmtwo.
The intensity ratio and electron density as a function of Galactic longitude  are shown in Fig. \ref{fig:ne_vs_l}.

\subsection{122~\um\ and 205~\um\ Intensity Comparison}
\label{122_205_comp}
\begin{figure}[!t]
    \centering
    \includegraphics[scale=0.9]{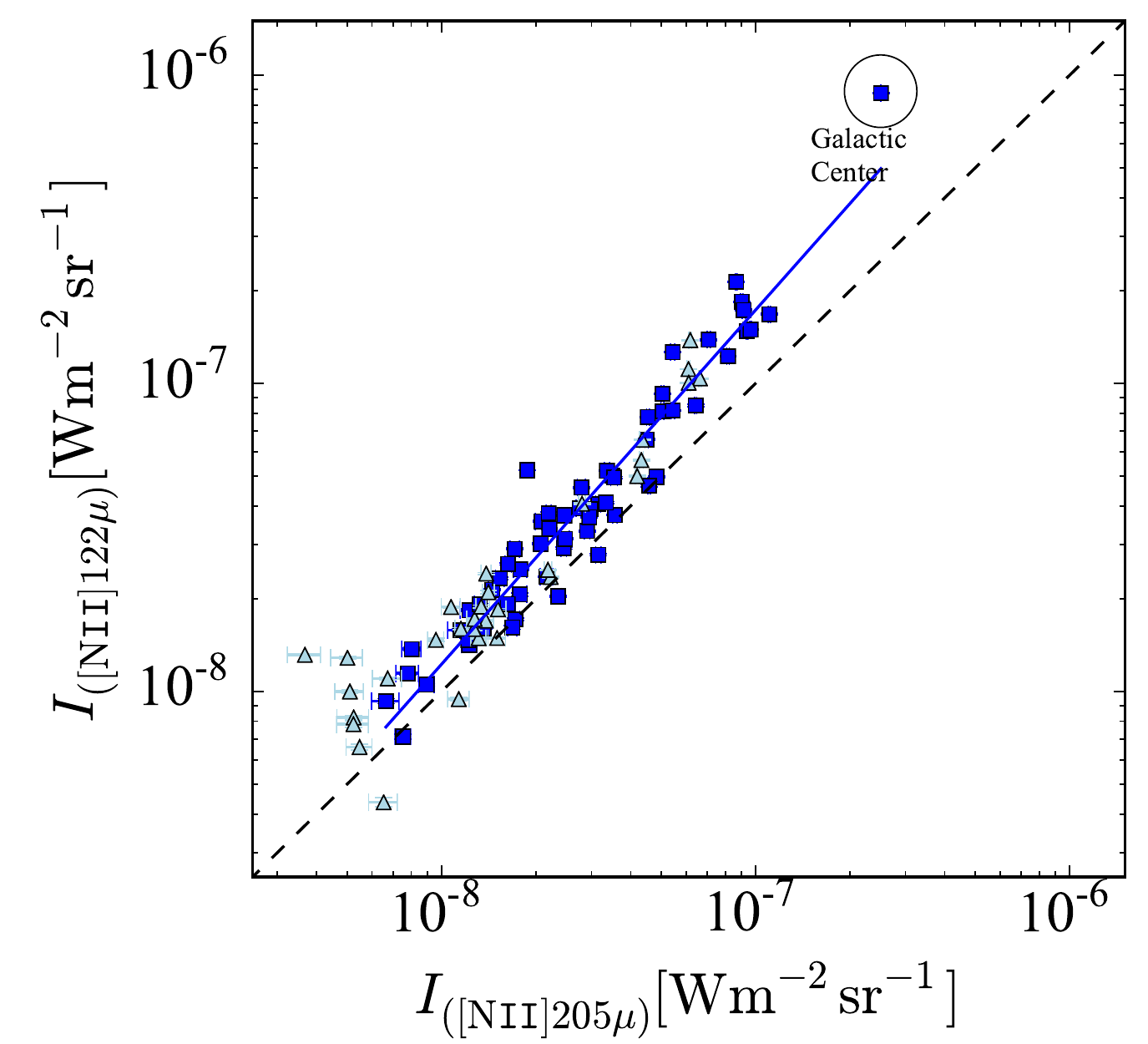}
    \caption{Comparison of the intensities of the  \NII\  of 122~\um\ and 205~\um\  lines.  The blue squares are the LoS in the range --30$\degr$$<$$l$$<$30$\degr$, and the triangles are the LoS positions in the region --60$\degr$$<$$l$$<$--30$\degr$ and 30$\degr$$<$$l$$<$60$\degr$. The black dashed line corresponds to equal intensities (1:1 ratio), and the blue line shows the fit to the data. The very high intensity point is the Galactic Center (G000.0+0.0).}
    \label{fig:Avg122vs205}
\end{figure}

Figure~\ref{fig:Avg122vs205} gives a comparison of the intensities of the two \NII\ lines for each pointing direction (the intensities of the 25 spaxels for each direction are averaged together).  We see an excellent correlation between the 122~\um\ and 205~\um\ \NII\ lines; the Pearson correlation coefficient is $r$=0.96 (7.4$\sigma$). Considering only the points with --30$\degr \leq\ l \leq\ 30\degr$  the best--fit relationship is given by $I({\rm 122}\mu m) \propto I({\rm 205}\mu m)^{1.15}$.

However, the slope is significantly greater than unity, which indicates that the regions having stronger N$^+$ emission also have higher electron densities.  This is plausibly a result of dominance of higher electron density regions even when combined with emission from extended low--density material, as discussed in \S \ref{model_results}.  The clouds with low values of $I$ (or of $N$) have  $I_{122}/I_{205}$ $\simeq$ 1.1 which corresponds to an electron density of 16 \cmthree, while the best fit curve for the most intense sources other than G000.00+0.0 (the Galactic Center) yields $I_{122}/I_{205}$ = 2, corresponding to $n(e)$ =  47 \cmthree.  The G000.00+0.0 observation deviates from the best fit relationship and stands out with $I_{122}/I_{205}$ = 3.5 and $n(e)$ = 118 \cmthree.

%
%

\subsection{Comparison of Derived Electron Densities with Previous Results}

The COBE FIRAS instrument covered both \NII\ fine structure transitions.  \citet{Bennett94} reported that if the instrumental line width is left as a free parameter, the average intensity ratio $<(I_{122}/I_{205})>$ = 1.6$\pm$0.2, which corresponds to 28~\cmthree~$\leq$~\ne\ $\leq$~44~\cmthree.  However, if the instrumental line width is fixed to that of the \CII\ line, the average ratio is unity, implying \ne\ = 12~\cmthree.  Given the very large beam size and the wide range of COBE results, we conclude only that the range of electron densities they derived may be consistent with those obtained here, and that even the COBE results yield much higher densities than those expected for the WIM.  

Almost contemporaneous with the COBE results were observations of \NII\ lines in the \HII\ region NGC333.6-0.2 by \citet{Colgan93}.  These data were obtained with a 45\arcsec\ beam size using the Kuiper Airborne Observatory (KAO).  Single spectra of each line were obtained, but the resolving power of $\simeq$1650 did not resolve the lines.  The electron density derived from the line ratio was 300 \cmthree.  This large value is consistent with modeling of the ionization structure traced by a variety of ions that were observed.  A more recent study including \NII\ was that of the Carina nebula by \citet{Oberst06,Oberst11}.  These authors combined ISO observations of the 122~\um\ line with 205~\um\ data from the AST/RO telescope outfitted with the SPIFI instrument.  The initial observations of a single point yielded $I_{122}/I_{205}$ = 1.5$\pm$0.64, with nominal $n(e)$ = 32 \cmthree, and range 7.2 \cmthree\ $\leq$~\ne\ $\leq$~60~\cmthree.  The later publication included electron densities at 27 positions spread within a region of 0.45$\degr$ in Galactic longitude and 0.15$\degr$ in Galactic latitude.  The average intensity ratio measured is $<(I_{122}/I_{205})>$ $\simeq$ 1.4  corresponding to \ne\ = 28 \cmthree, but including a range extending from \ne\ = 3 \cmthree\ to 122~\cmthree.  The range and mean values of \ne\ agree well with those found in the present study.  However, since Carina is an \HII\ region powered by a large cluster of O--stars, extended ionized gas of this density is not surprising.  Finally, \citet{Langer15} used observations of the 205~\um\ line made with the GREAT instrument on SOFIA, together with a model for the geometry and parameters of the ionized layer from \CII, to derive densities between 5 \cmthree\ and 21 \cmthree\ for two lines of sight near the edge of the Central Molecular Zone (CMZ).  These again are reasonably consistent with the densities derived here.

\section{Discussion}
\label{sec:discussion}

\subsection{\NII\ and \CII\ Emission} 

\begin{figure}[!t]
    \centering
    \includegraphics[scale=0.8]{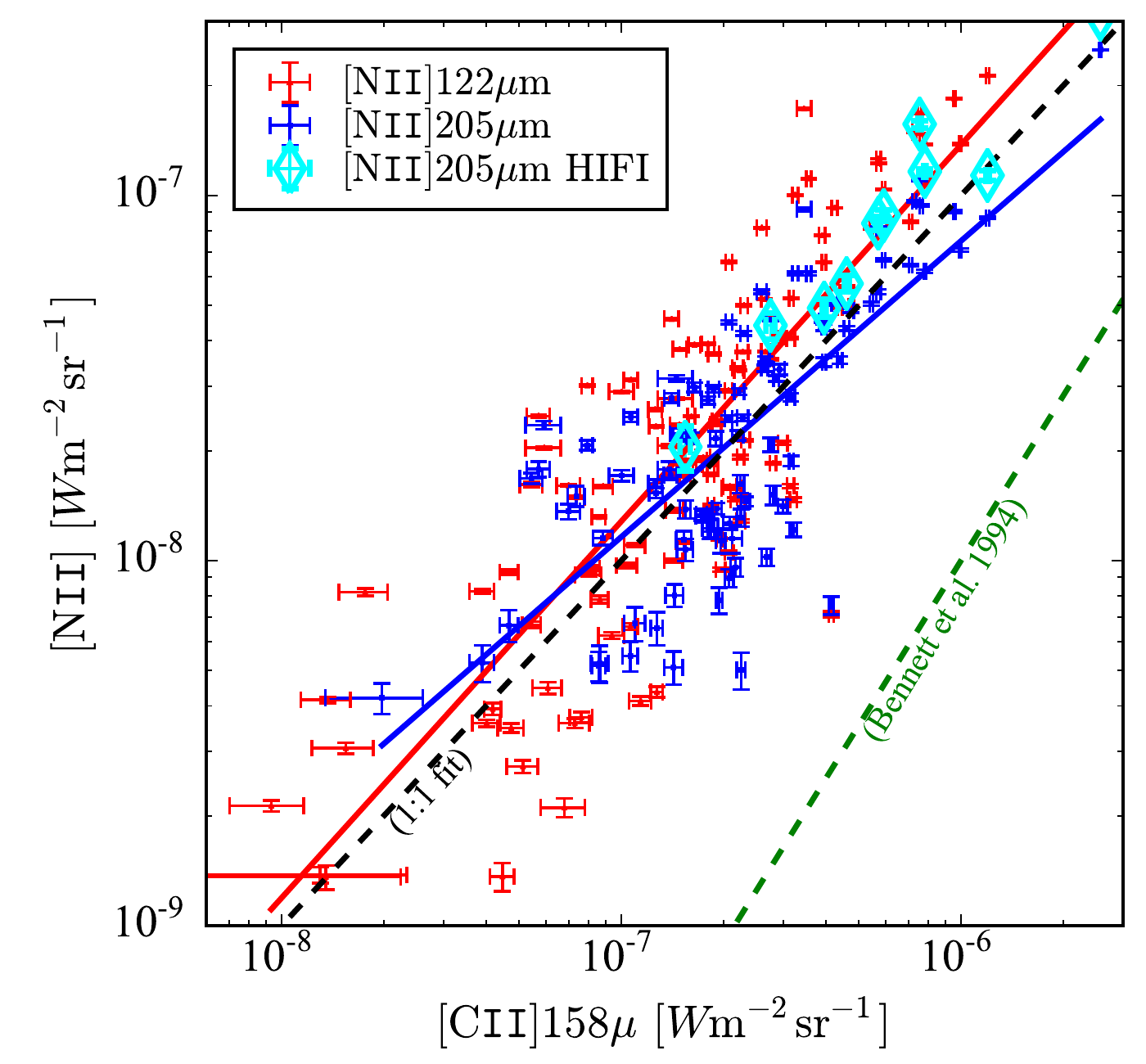}
    \caption{ Intensities of \NII\  122~\um\ and 205~\um\ transitions compared with those of \CII\  158~\um. The red and blue solid lines are power law fits of the corresponding \NII\ lines versus \CII.  The black dashed line is  a slope of unity, indicating that both \NII\ transitions and also \NII\ measured with HIFI as well as PACS are linearly correlated with the \CII\ emission.  The green dashed line shows a slope of 1.5, as derived by \citet{Bennett94}.}
    \label{fig:NplusCplus}
\end{figure}

A possible clue to the origin of the \nplus\ emission is comparison with that of \cplus.  The latter can come from a wide variety of sources since this ion can be produced by photons of energy $>$ 11.26 eV, and is thus relatively widespread.  We compare fluxes for the two \NII\ transitions observed here with those of \CII\ from the GOT C+ survey in Fig.~\ref{fig:NplusCplus}.
The \NII\ 122~\um, 205~\um\ (PACS), and 205~\um\ (HIFI) data are compared separately to the \CII\ (HIFI) data in the figure.  The  205~\um\ data have a best--fit slope less than unity and the 122~\um\ data have a slope greater than unity by a comparable amount.  The HIFI data are fit by a slope very close to unity.  All three slopes that we derive are significantly closer to unity than the value of 1.5 found by \citet{Bennett94}, making their suggested explanation of \NII\ coming from a volume while \CII\ comes from its surface less compelling.

Given that we have the electron density and the \nplus\ column density in the ionized region responsible for the \NII\ emission, we can readily compare this with the \cplus\ column density from the GOT C+ project.  For the \cplus, we start with Equation~26 of \citet{Goldsmith12}, which gives the integrated antenna temperature produced by a given quantity of ionized carbon with a given collision rate.  Since these {\it Herschel} HIFI observations were taken with a diffraction--limited system, we can convert the integrated antenna temperature to intensity using the relationship 
$\int T_{\rm A}dv$ [K km s$^{-1}$] = 10$^{-3}(\lambda^3/2k)I [{\rm W}m^{-2}sr^{-1}].$ 
The resulting expression for the column density of \cplus\  excited by electrons at 8000 K at density $n(e)$ that produce intensity $I$ is
\begin{equation}
N({\rm C}^+) [{\rm cm^{-2}}] = 4.14\times10^{23}[1 + 0.506(1 + \frac{46.1}{n(e)})]I(158\mu m) [{\rm Wm^{-2}sr^{-1}}]~.
\end{equation}

\begin{figure}[!t]
    \centering
    \includegraphics[scale=0.8]{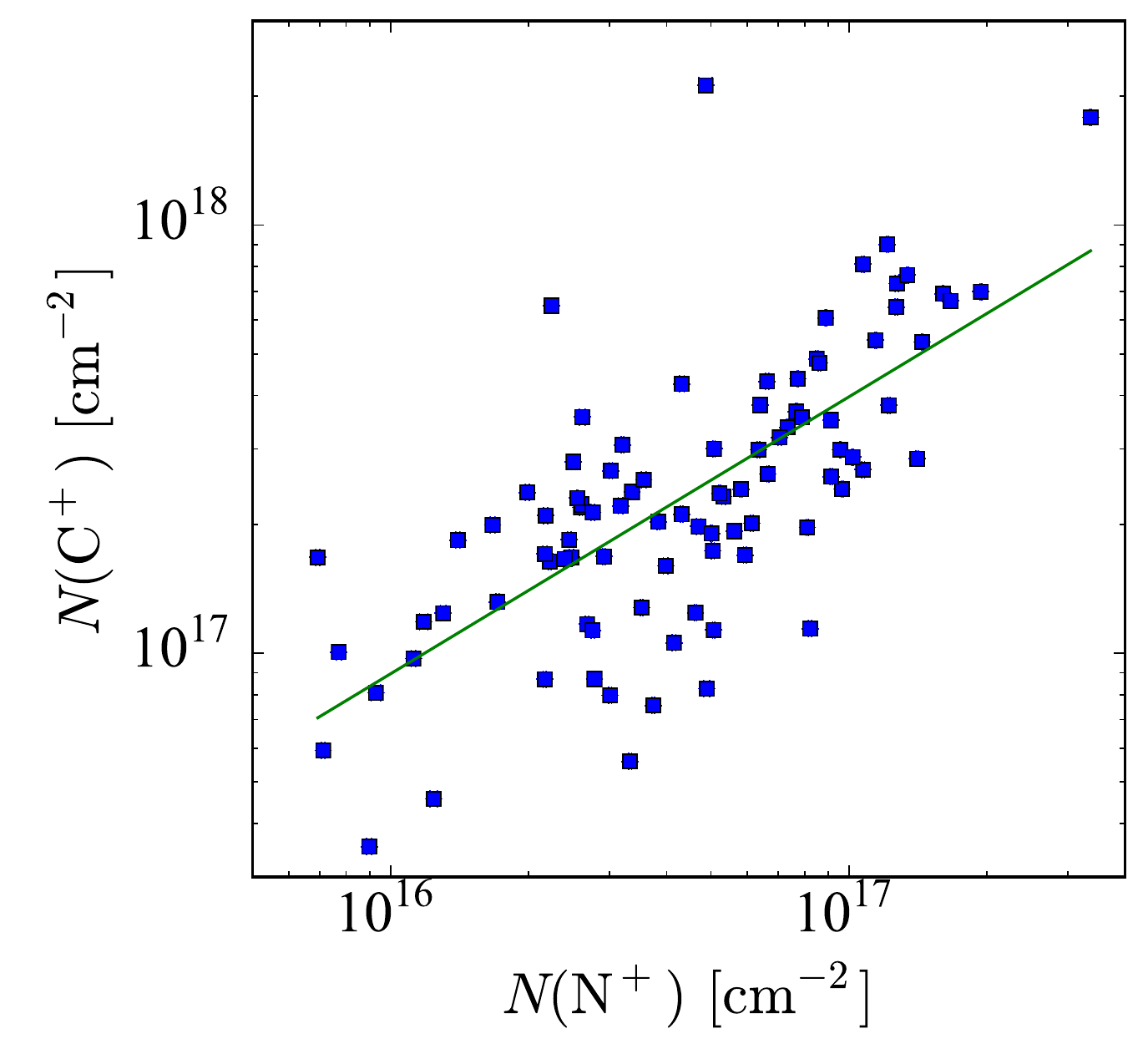}
	\caption{ \cplus\ column density that would be implied by GOT C+ data if all ionized carbon emission were from the same ionized gas as responsible for the \nplus\ emission, plotted as a function of the \nplus\ column density in the same LoS.  The green line is the best fit linear relationship and indicates that with this assumption, $N$(\cplus) $\propto$ $N$(\nplus)$^{0.65}$.  }
	\label{fig:NNII-NCII}
\end{figure}

It is straightforward to calculate the column density  of \cplus\ that we infer from the GOT C+ data if we assume that all of the emission were produced in the same regions responsible for the \nplus.  The comparison of the resulting column densities is shown in Fig.~\ref{fig:NNII-NCII}.
The column density ratio is close to 10 for directions with $N$(\nplus) $\simeq$ 10$^{16}$ cm$^{-2}$, and close to 5 for directions with an order of magnitude larger \nplus\ column density.  The slope of the best fit linear relationship is 0.65, but some of the highest column density directions (including G000.0+0.0) have a column density ratio again close to 10.
We can compare the ratio of \cplus\ to \nplus\ column densities with the expected C/N abundance ratio.  As discussed in the references given in Section \ref{IBL}, at a nominal adopted Galactocentric distance of 1.5 kpc, the elemental abundance ratio is $\simeq$ 2.9 as compared to 4.0 in the Sun.  

The clear implication of the measurements shown in Fig.~\ref{fig:NNII-NCII} is that instead of a factor $\simeq$3 greater \cplus\ column density compared to that of \nplus\ as would be the case if all carbon were \cplus\ in the same regions where nitrogen is \nplus, we see factor of 5 to 10.   While the exact fraction depends on the Galactocentric distance and elemental abundance of the emitting regions, we conclude that between 1/3 and 1/2 of the total \CII\ emission is produced in the ionized gas.  Based on where we have detected \NII\ emission, we have been considering this gas to be coming from the inner few kpc of the Milky Way.  There, \citet{Pineda13} do see a significant fraction of emission from ionized gas, so the fraction derived here, though somewhat larger, is not in serious disagreement. Further detailed modeling as well as velocity-resolved observations will be required to more accurately determine the \cplus\ fractional emission from ionized gas.

\subsection{Electron Density -- \nplus\ Column Density Relationship}
\label{ne-N}
\begin{figure}[!t]
	\centering
    	\includegraphics[scale=1.0]{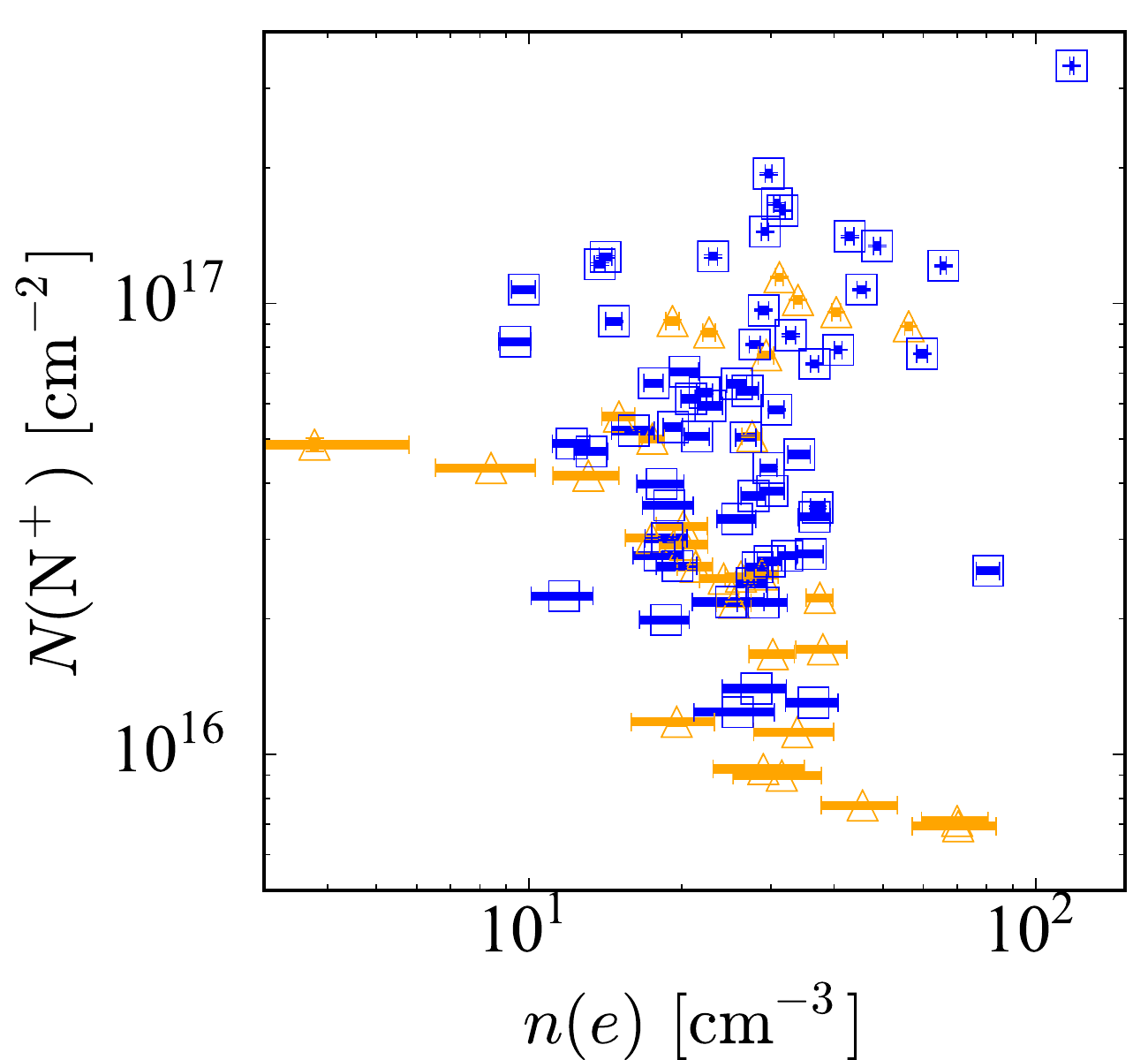}
	\caption{\nplus\ column density compared to the electron density derived from the \NII\ fine structure line ratio.  The dark blue squares are the LoS in the range --30$\degr$$<$$l$$<$30$\degr$, and the orange triangles denote the LoS positions in the region --60$\degr$$<$$l$$<$--30$\degr$ and 30$\degr$$<$$l$$<$60$\degr$. }
	\label{fig:ne-N}
\end{figure}

Figure~\ref{fig:ne-N} shows a comparison of the electron density derived from the \NII\ fine structure lines with the \nplus\ column density.
While there is no global trend, there is a suggestion that for $N$(\nplus)  $\geq$ 3$\times$10$^{16}$ \cmthree, there is a positive correlation of column density with electron density, while for lower column densities, there is no clear trend visible.    There are three points having very low column densities and high electron densities; in these directions the intensities were sufficient that a detection was made even without averaging all spaxels together.  The high ratio of 122~\um\ intensity to 205~$\mu$m intensity indicates a high electron density, but the very low values of the intensities then result in unusually low column densities.  There does not seem to be a reason for rejecting these data, although the result is clearly somewhat unusual.

\subsection{Origin of the Observed \NII\ Emission}
\label{origins}

\subsubsection{Warm Ionized Medium (WIM)}
\label{WIM}
Both \NII\ fine structure transitions were detected over much of the inner Galaxy, with characteristic column densities $N({\rm N^+})$ = 4$\times$10$^{16}$ \cmtwo\ and electron densities \ne\ = 30~\cmthree.  While the column densities do not differ radically from those expected in the WIM \citep [see e.g.,][]{Persson14}, the electron densities are two to three orders of magnitude greater than those from the ``standard'' WIM; see e.g., \citet{Reynolds91}, \citet{Hill08}, and \citet{Ferriere01} and references therein.  The thermal pressure implied by the presence of ionized gas at 8000 K \citep{Haffner99} with a density $\simeq$ 30 \cmthree\ is one to two orders of magnitude higher than that of other well--accepted ISM phases.  The results overall suggest a different source for the observed \NII\ emission.  \citet{Reynolds01} suggested, based on H$\alpha$ observations away from the Galactic Plane, that the temperature of the WIM is substantially warmer, $\simeq$ 10$^4$ K, than generally assumed. 
Both the \cite{Tayal11} as well as the \citet{Hudson04} collision rate coefficients have a weak inverse dependence on the kinetic temperature.  An increase in $T_{\rm k}$ from 8000 K to 10000 K would decrease the deexcitation rate coefficients by $\simeq$ 10\%, resulting in this same fractional increase in the derived electron density, making it even more discrepant with that of the standard WIM.

The WIM has long been recognized as not having a single density.  In their Galactic model, \citet{Taylor93} included three components having densities between 0.02 and 0.1 \cmthree.  A revision to this model by \citet{Heiles96} added a low volume filling factor ($\phi_v$ $\simeq$ 0.01) but relatively high density, \ne\ = 5 \cmthree\ component to the spiral arms.  This was one of the observational points included by \citet{Berkhuijsen98} in a compilation of WIM component electron densities and filling factors, that suggested that $\phi_v$ $\simeq$ $n(e)^{-1}$.  \citet{Mitra04} found a very similar density--volume filling factor relationship,  $\phi_v$ =  =0.0184$n(e)^{-1.07}$ for ionized gas within 3 kpc of the sun.  A modest extrapolation could include regions with \ne\ $\simeq$ 30 \cmthree, which would be anticipated to have $\phi_v$ somewhat greater than 0.001.  If such regions are a result of the turbulent structure of the WIM they may not be in photoionization equilibrium, but rather could be short--lived, as the recombination time is only $\simeq$ 4000 yr \citep{Draine11}. These might be related in some way to the highly overpressured regions within diffuse (atomic and molecular) clouds suggested by \cplus\ UV line ratios  \citep{Jenkins11}.

\subsubsection{Ionized Boundary Layers (IBL)}
\label{IBL}
\begin{figure}[!t]
    \centering
    \includegraphics[scale=0.8]{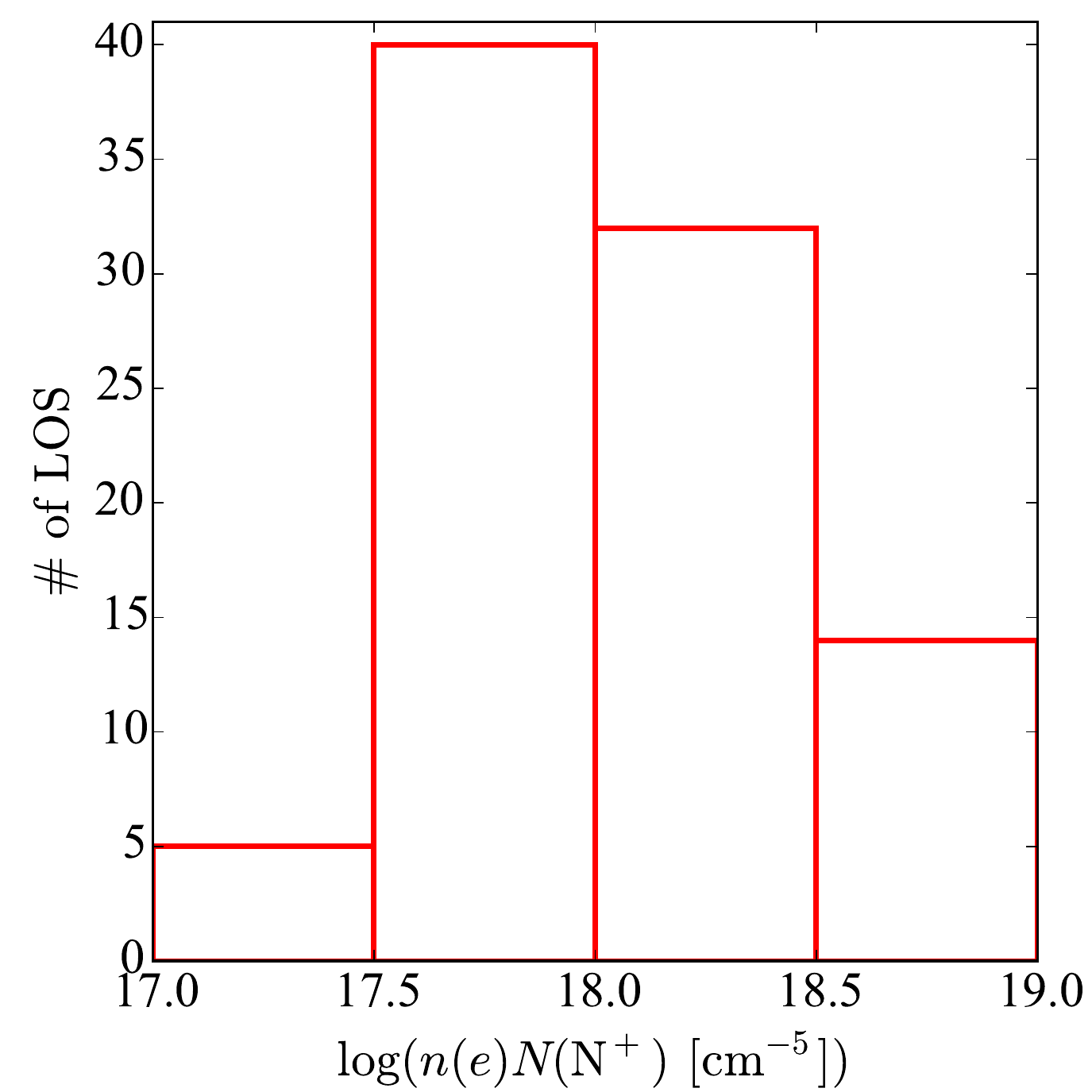}
    \caption{\small Distribution of observed N$^+$ column density multiplied by the derived electron density}
    \label{fig:n(e)N}
\end{figure}

A possible source for the \NII\ emission is the dense ionized boundary layers (or IBL) \citep {Bennett94}, that plausibly border any dense portion of the ISM subjected to a flux of photons from massive young stars.  Given the significant unknowns in modeling, we adopt a very simple procedure for estimating the parameters of such a model.  We first have to consider that a typical \NII\ spectrum observed with PACS is the integral of a number of distinct kinematic components, as indicated by our limited HIFI results (Fig.~\ref{fig:HIFINIICIIspectra}).  The number of components ranges from one to five, and taking each to have two surfaces exposed to an external radiation field, we find that the average number of fields along a single line of sight is 5.  We show in Fig. \ref{fig:n(e)N}, the distribution of the product $n(e)N({\rm N}^+)$, which has a characteristic value of 1$\times$10$^{18}$ cm$^{-5}$.  It is reasonable to assume that this quantity is divided among the surfaces, so that the characteristic quantity {\it per surface} is approximately 2$\times$10$^{17}$ cm$^{-5}$.

A major question is the photon flux necessary to sustain the required column density of H$^+$.  The latter is the the critical parameter since in a hot hydrogen plasma, charge exchange with nitrogen, H$^+$ + N $\rightarrow$ H + N$^+$, will assure that the nitrogen is essentially completely ionized \citep{Lin05,Langer15}.  To calculate the column of ionized hydrogen, we adopt a model described by \citet{Davidson79}, which assumes that all of the incident ionizing photons are absorbed in a length $L$, and assumes that this flux $F$ (photons cm$^{-2}$s$^{-1}$) is balanced by the rate of hydrogen ion recombinations within the column.  Assuming that the region has a uniform density and is completely ionized, $n(e)$ = $N({\rm H^+})$ = $n$.  With ionized hydrogen recombination coefficient $\alpha$, we then have
\beq
F = \alpha n^2 L ~.
\eeq
With fractional abundance of ionized nitrogen $X({\rm N^+})$, and N$^+$ column density $N({\rm N^+}) = X({\rm N^+})nL$, we find that
\beq
F = \frac{\alpha}{X({\rm N^+})} n(e) N({\rm N^+}) \lp
\eeq

We start with a local fractional abundance of nitrogen relative to hydrogen of 6.76$\times$10$^{-5}$ \citep{Asplund09}, but scaling this local value by an abundance gradient --0.07 dex/kpc suggested by \citet{Shaver83} and \citet{Rolleston00}, we find for a representative Galactocentric distance of 1.5 kpc that $X(\rm N^+)$ = 2.9$\times$10$^{-4}$.  With the recombination coefficient from \citet{McElroy13} $\sim$ 3$\times$10$^{-13}$ cm$^{-3}$s$^{-1}$, we obtain
\beq
F \simeq\ 10^{-9}n(e) N({\rm N^+}) \lp
\label{F_num}
\eeq
With $n(e)N({\rm N}^+)$ equal to 2$\times$10$^{17}$ cm$^{-5}$ per surface, Equation~\ref{F_num} indicates that a flux of 2$\times$10$^8$ hydrogen--ionizing photons cm$^{-2}$ s$^{-1}$ is required. 
We may consider a single O6 star, that produces 4$\sim$10$^{49}$ hydrogen--ionizing photons per second as a potential source of radiation. Assuming that there is no absorption between the star and the surface of the IBL, we find that the distance to the star can be as great as $\simeq$ 14 pc and still provide a sufficient flux of ionizing photons.   A cluster of massive young stars producing 100 times greater ionizing flux could be as distant as 140 pc.
The IBL could thus contribute if these regions were sufficiently numerous in the central portion of the Galaxy.

\subsubsection{HII Region Envelopes}
\label{envelopes}

Typical \HII\ regions have densities considerably in excess of those found characteristic of the \NII\ emission, but there is a long history of low density \HII\ regions and \HII\ region envelopes that is likely relevant to the issue of the origin of the \nplus\ fine structure line emission.  \citet{Shaver76} concluded that cm--wavelength recombination lines required relatively high electron densities, 5 to 10 \cmthree\ in the central portion of the Milky Way.  Modeling of star formation and diffuse infrared radiation led \citet{Mezger78} to infer the presence of ``Extended Low Density" (ELD) \HII\ regions with effective electron density $\simeq$ 3 \cmthree.  \citet{Anantharamaiah85, Anantharamaiah86} compared low--frequency and high--frequency recombination lines, and from their relative intensities, concluded that the low--density envelopes of \HII\ regions, with electron densities between 1 and 10 \cmthree\ were the source of the observed emission.  Based on modeling pulsar emission measure data, \citet{McKee97} determined that typical \HII\ region envelopes in the Milky Way are characterized by an electron density $n(e)$ = 2.7 \cmthree.  Ionized gas in the 1 to 10 \cmthree\ was required to reproduce the recombination line results of \citet{Roshi01}.  These authors also concluded that a considerable fraction of the \CII\ and \NII\ emission observed by the {\it COBE} satellite also originates in these \HII\ region envelopes.

It is certainly plausible that there is an intermediate-density region between these envelopes and the far denser \HII\ regions themselves.  Through a combination of the relatively smaller angular size and consequent beam dilution, we do not typically see densities characteristic of ``young'' \HII\ regions ($\geq$ 100 \cmthree), but due to the larger excitation rate due to higher electron density, we typically do not derive densities as low as those characteristic of the envelopes discussed above.  Figure~\ref{fig:ne-N} in Section \ref{ne-N} shows that the LoS directions with the largest \nplus\ column densities have some of the highest electron densities.  This might be what one would expect for lines of sight traversing the portions of \HII\ regions with high emission measure ($n(e)N(e)$) which also tend to be the more compact \HII\ regions or portions thereof.  This relationship between electron density and \nplus\ column density (or that suggested by the relationship between the intensites of the two \NII\ transitions shown in Fig.~\ref{fig:Avg122vs205}) would not be produced in any obvious way by WIM or IBL emission, and thus favors the \HII\ region envelopes as the source of the observed \NII\ fine structure line emission.

\subsection{Uncertainties in the Density Determination}

\subsubsection{Combination of Regions Having Different Densities}

One might be concerned that when the emission from the traditional, very low--density WIM is added to the emission from a ``high density'' region, the resulting density determination would be erroneous.  As discussed in Section \ref{multiple}, the addition of a low density region has a limited effect since the \nplus\ there is almost entirely in the ground state.  Given the measured column density of the WIM seen in absorption (Section \ref{tau}), the ratio of the column density in the low density region to that in the high density region is certainly less than 10, and the perturbation is very small.  Increasing the derived electron densities by 10\% would bound the magnitude of the effect.  

\subsubsection{Galactic Electron Temperature Gradient}
\label{gradient}

A variety of observations have long indicated that the heavy element abundance gradient in the Milky Way produces a gradient of the cooling rate of ionized gas in \HII\ regions.  The increased metal abundance towards the Galactic Center (e.g. Section \ref{IBL}) results in a lower electron temperature as one moves inward.  A recent study by \citet{Quireza06} has results that depend on the sample of \HII\ regions employed.  We adopt their Sample F that gives $T_e ({\rm K})$ = 5730 + 268 $R$(kpc).  The resulting electron temperature at a Galactocentric radius of 1.5 kpc is 6130 K, compared to 8000 K at 8.5 kpc.  Following the discussion in Section \ref{WIM}, the lower kinetic temperature results in an increase of $\simeq$ 11\% in the rates, and a comparable (or slightly smaller) decrease in the electron density derived from the \NII\ line ratio.

\subsubsection{Optical Pumping of Fine Structure Levels}
\label{pumping}

Absorption of infrared, optical, or UV radiation followed by spontaneous emission back to the ground electronic state can populate the excited fine structure lines of \nplus.  This mechanism has been considered in detail by \citet{Flannery79}, who conclude that this process is unlikely to be important in \HII\ regions either when densities in the ionized gas are low and the radius of the Str\"{o}mgren sphere is large or when the \HII\ region is compact and dense.  We can gain some insight into the plausible significance of pumping in various situations by simply comparing the upwards ``pump'' rate produced by photon absorption (which is approximately the rate at which population can be transferred from a lower to an upper fine structure level) to the collision rate we infer from our data.  If the pump rate is much smaller than the collision rate among the fine structure levels required to produce the observed intensity ratio $I_{122}/I_{205}$, we can safely ignore the pumping process.  

As an example, we consider the 1083.99 \AA\ wavelength UV transition from the $^3$P$_0$ level to the $^3$D$_1$$^0$ level, which has a spontaneous decay rate of $\simeq$2$\times$10$^8$ s$^{-1}$. For this transition, we derive an integrated absorption cross section $\int \sigma(\nu)d\nu$ = 0.0031 cm$^{-2}$Hz.  Since radiation fields are generally expressed in terms of $\nu U_{\nu}$ having units erg cm$^{-3}$, we can express the pump rate as $R_{\rm p} = (c/h\nu^2)[\nu U_{\nu}]\int \sigma(\nu)d\nu$. The resulting expression is $R_{\rm p} = 1835[\nu U_{\nu}]$.  Using the information given in Chapter 12 of \citet{Draine11}, we find that at a frequency of $\simeq$3$\times$10$^{15}$ Hz, in the solar neighborhood, $\nu U_{\nu}$ $\simeq$ 7$\times$10$^{-14}$ erg cm$^{-3}$, and $R_{\rm p}$ = 1.3$\times$10$^{-10}$ s$^{-1}$, which is less than 1/100 of the derived collision rate.  A similar result follows from the radiation field in a typical \HI\ cloud (Draine Figure 12.1).  At the boundary of a \HII\ region within 0.2 pc of exciting O8V star, it is possible for $\nu U{_\nu}$ to exceed 10$^{-9}$ erg cm$^{-3}$ (Draine Figure 12.3), at which point the pumping rate becomes comparable to the inferred collision rate.  It is thus possible that in extreme cases, optical pumping could be significant but more detailed calculations are required to assess its effect as suggested by the complexities discussed by \citet{Flannery79}.  

\section{Conclusions}
\label{sec:conclusions}

We have presented the initial results from the first high angular resolution Galactic plane survey of the two \NII\ far--infrared fine structure transitions.  We used the PACS instrument on the {\it Herschel} Space Observatory to observe 149 positions in both the 122~$\mu$m and 205~$\mu$m transitions, and the HIFI instrument to observe the lower--frequency line at 10 of these positions in the Galactic Plane.  Our results can be summarized as follows.

We have corrected for the long--wavelength leak affecting the PACS data and a comparison of the 205~$\mu$m intensities with the corresponding HIFI data indicates agreement within approximately 20\%.  The central spaxel detected emission from 94 telescope pointing directions in the 122~\um\ line and 59 directions in the 205~$\mu$m line, but when all 25 spaxels were averaged together, the number of detections increased to 116 at 122~\um\ and 96 at 205~$\mu$m.  We detect emission in both transitions, with the detections being largely towards the center of the Galaxy, with --60\degr $\leq$ $l$ $\leq$ +60\degr.  The mean intensity of both \nplus\ fine structure lines is $\simeq$ 2$\times$10$^{-8}$ $W$m$^{-2}$sr$^{-1}$.  The \NII\ emission is relatively uniform over the 25 spaxels observed in each telescope pointing, with $<I_{\rm rms}/ \overline{I} >$  equal to 0.25.  \NII\ emission is definitely not produced by pointlike sources.

We have carried out analytic and numerical modeling of the collisional excitation of the \nplus\ fine structure transitions by electrons, since we expect this ion to be present only in regions in which hydrogen is essentially fully ionized.  Using collision rate coefficients from the literature, we show that the ratio of the 122~$\mu$m to 205~$\mu$m intensity is a good tracer of electron density for 10 \cmthree\ $\leq$ $n(e)$ $\leq$ 1000 \cmthree, with little concern about optical depth effects and radiative trapping.  For very low electron densities, the fractional population of the upper two fine structure levels is very small. In consequence, the ratio of the two \NII\ fine structure lines produced in a region with $n(e)$ $\geq$ 10 \cmthree\ is hardly affected by material having $n(e)$ $<$ 1 \cmthree\ unless the column density of the latter approaches 100 times that of the former.  

We find electron densities ranging from $\simeq$10 \cmthree\ to $\simeq$100 \cmthree, with a mean value $<n(e)>$ = 33 \cmthree.  The total  \nplus\ column densities range from 3$\times$10$^{15}$ \cmtwo\ to 3$\times$10$^{17}$ \cmtwo, with a mean value $<N({\rm N}^+)>$ = 4.7$\times$10$^{16}$ \cmtwo.  However, HIFI data indicate that $\simeq$ 5 discrete kinematic features make up a typical \NII\ emission profile in the inner Galaxy.  The PACS data (which integrate the entire line) thus indicate a typical ionized layer column density of $\simeq$10$^{16}$ \cmtwo.  The electron densities are consistent with those found by COBE survey with a 7\degr\ beam as well as with observations of individual clouds made with beam sizes comparable to those employed here.  There is no obvious trend of $n(e)$ or of $N({\rm N}^+)$ with Galactic longitude.  

Comparison of the HIFI \NII\ 205~$\mu$m spectra with  \CII\ 158~$\mu$m spectra from the GOT C+ project reveals that the two fine structure lines emit at the same velocities. There is no clear instance of \CII\ without \NII\ or vice--versa.  However, the ratio of the antenna temperatures of the two lines varies significantly.  We find a good correlation between the \NII\ and \CII\ intensities, with a slope close to unity, which differs significantly from the slope of 1.5 found from COBE data.  If we assume that the \cplus\ emission is produced in the same ionized gas as the \nplus, we can compare the column density of ionized carbon from the GOT C+ data and find that for $N$(\nplus) = 1$\times$10$^{16}$ \cmtwo, $N$(\cplus)/$N$(\nplus) $\simeq$ 10, but the ratio has a slope of 0.65 extending upwards a factor $\simeq$ 10 in column density.  With simple assumptions about the location of the emission and Galactic gradient in elemental abundances, we find that between 1/3 and 1/2 of the total \CII\ emission is produced by fully ionized gas.  The electron density and \nplus\ column density do not have a clear correlation for $N$(\nplus) $\leq$ 3$\times$10$^{16}$ \cmtwo, but for $N$(\nplus) greater than this, there is a suggestion that $N$(\nplus) and $n(e)$ increase together.

We have considered several different scenarios to determine the origin  of the \NII\ emission with the characteristics observed.  The standard WIM has too--low a density by a factor of $\simeq$ 100 or greater to explain the derived $n(e)$, but it is possible that we are seeing high--density condensations which are most likely transient structures in the generally very diffuse ionized gas.  UV radiation from a massive young star or cluster can produce a layer of ionized gas (IBL) before the ionizing photons are absorbed and we move into the  hot, initially atomic photon dominated region (PDR).  The column density of ionized gas at the density observed demands a relatively intense radiation field and thus a nearby source of photons.  This scenario is attractive in that it readily explains the additional \cplus\ as being in the PDR but there is a question whether the radiation from exciting stars is sufficient to explain the \NII\ emission's uniformity and extension throughout the central portion of the Milky Way.  A third possibility is that we are seeing the relatively low density envelopes of \HII\ regions.  The $n(e)$ we derive are intermediate between the very Extended Low--Density (ELD \HII) fully ionized gas that has been identified from free--free and low frequency radio recombination line studies, and the densities for \HII\ regions themselves.  The ELD gas is thought to fill a significant fraction of the volume of the inner Galaxy, and the material responsible for the \NII\ emission, with a density one order of magnitude greater, could plausibly have a sufficient volume filling factor to reproduce the extended \NII\ emission we have observed.
\clearpage

\begin{acknowledgements}

The authors are grateful to Steve Lord, Dario Fadda, and Roberta Paladini of the 
NASA {\it Herschel} Science Center for assistance with the PACS data reduction. 
Gordon Stacey and Gary Melnick provided useful information on issues of calibration.
Moshe Elitzur, Neal Evans, Carl Heiles, and Joe Lazio provided valuable inputs on a variety of topics.  Roberta Paladini shared useful information on low density components of the ionized interstellar medium.
Carina Persson and John Black gave us valuable information on \NII\ lines and in particular on optical pumping.
We thank the refereee, Gordon Stacey, for corrections and suggestions that significantly improved this paper.
The data used herein were obtained with the {\it Herschel} Space Observatory using the PACS and HIFI instruments.  {\it Herschel} is an ESA cornerstone mission with major NASA participation.
This research was conducted at the Jet Propulsion Laboratory, which is operated by the California 
Institute of Technology under contract with the National Aeronautics and Space Administration (NASA). 
\copyright\ 2015 California Institute of Technology. \\

\end{acknowledgements}

\bibliography{bibdata}


\end{document}